\documentclass{article}
\usepackage{arxiv}
\usepackage{enumitem}
\usepackage[authoryear, sort&compress, round]{natbib}
\usepackage{bbm}
\usepackage{algorithm}
\usepackage{algpseudocode}
\usepackage{enumitem}
\setlist{nosep}
\usepackage{graphicx}
\usepackage{tabularx}
\usepackage{booktabs}
\usepackage{subcaption}
\usepackage{colortbl}
\usepackage{hyperref}
\usepackage{verbatim}
\usepackage{float}
\usepackage{amsmath}
\usepackage{cleveref}
\usepackage{amssymb}
\usepackage{pifont}

\usepackage{wrapfig}
\usepackage{makecell}
\usepackage{multirow}
\usepackage{fancyvrb}
\usepackage[normalem]{ulem}
\usepackage{url}
\usepackage{tcolorbox}
\usepackage{diagbox}
\usepackage{CJKutf8}
\usepackage{dashrule}
\tcbuselibrary{most}
\usepackage{xcolor}
\usepackage{arydshln}
\usepackage{calligra}
\usepackage{cancel} 
\usepackage[dvipsnames]{xcolor}

\usepackage{makecell}
\usepackage{array}
\usepackage{ragged2e}
\tcbuselibrary{breakable} 
\usepackage{minted}
\usepackage{soul}

\newcolumntype{Y}{>{\RaggedRight\arraybackslash}X}
\newcolumntype{Z}{>{\RaggedRight\arraybackslash}p{\dimexpr 5\linewidth/11-2\tabcolsep}}
\usepackage{threeparttable} 
\usepackage{calc}         

\usepackage{fontawesome5}

\newtcolorbox{mybox}[2][]
  {colback = black!5!white, colframe = black!75!black, fonttitle = \bfseries,
    colbacktitle = black!100!black, enhanced, before upper={\fontsize{10}{11}\obeyspaces\obeylines\selectfont}, fontupper=\selectfont,
    attach boxed title to top left={yshift=-2.2mm,xshift=2mm},
    title=#2,#1}

\newtcolorbox{promptbox}[3][]{
colback=black!5!white,
arc=5pt, 
boxrule=0.5pt,
fonttitle=\bfseries,
title=#3, 
before upper={\fontsize{8}{11}\obeyspaces\obeylines\selectfont}, 
fontupper=\selectfont,
colframe=#2,
label=#3,
}

\newtcolorbox{caseStudyBox}[2][]{
    breakable, 
    title=#2,  
    colback=gray!5!white, 
    colframe=gray!60!black, 
    fonttitle=\bfseries, 
    coltitle=black,
    attach boxed title to top left={yshift=-2mm, xshift=4mm}, 
    boxed title style={
        colback=white,
        colframe=gray!60!black,
    },
    #1 
}

\definecolor{lightgreen}{RGB}{145, 204, 117}
\definecolor{lightblue}{RGB}{173, 216, 230}
\definecolor{techblue}{RGB}{70, 130, 180}
\definecolor{academicred}{RGB}{178, 34, 34}
\definecolor{coolgray}{RGB}{128, 128, 128}

\newlist{questions}{enumerate}{2}
\setlist[questions,1]{label=RQ\arabic*.,ref=RQ\arabic*}
\setlist[questions,2]{label=(\alph*),ref=\thequestionsi(\alph*)}

\title{\includegraphics[width=0.03\textwidth]{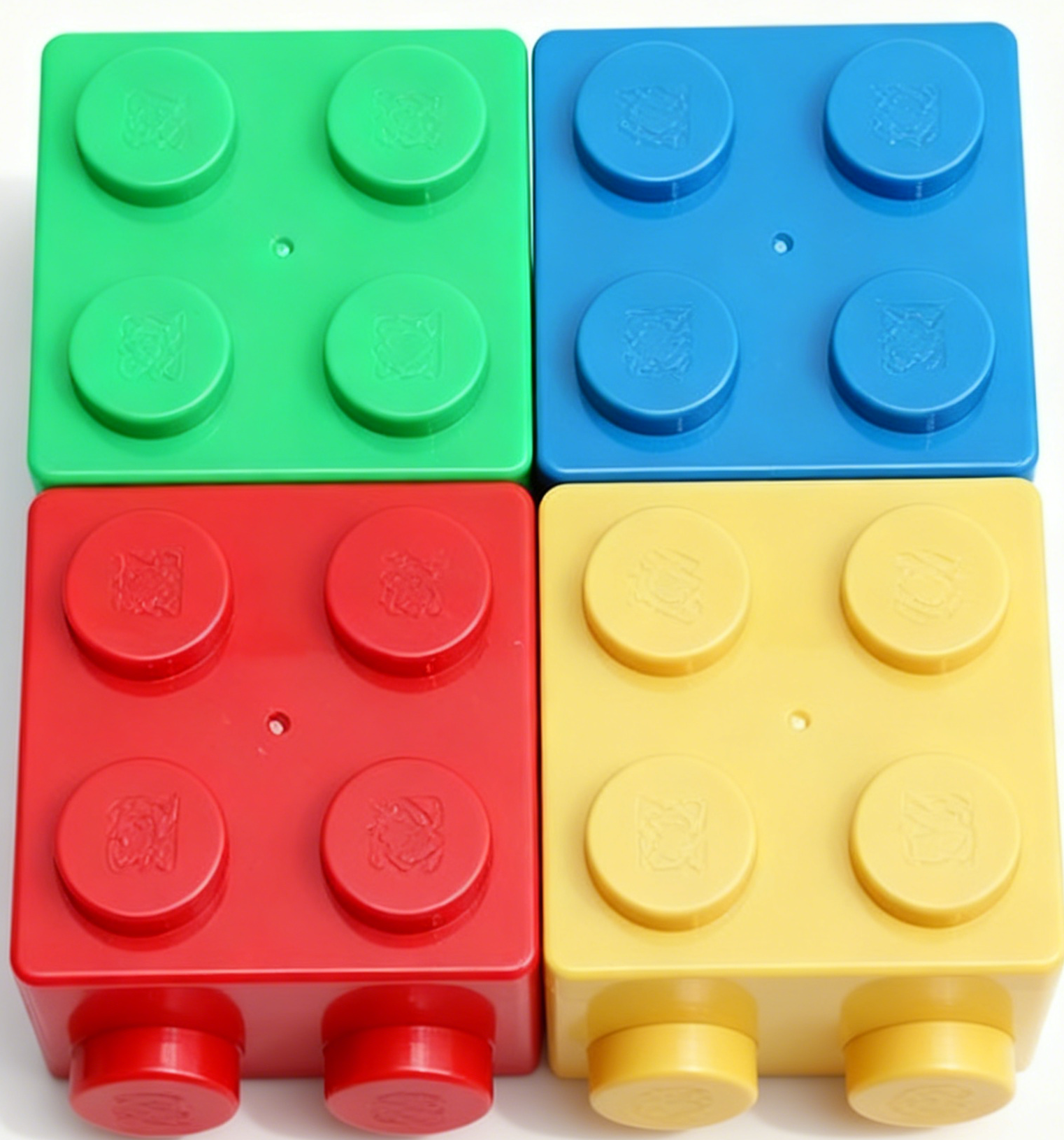} SWE-Lego: Pushing the Limits of Supervised Fine-tuning for Software Issue Resolving}

\author{
Chaofan Tao$^{*}$\textsuperscript{1}
Jierun Chen$^{*}$\textsuperscript{1}
Yuxin Jiang$^{*}$\textsuperscript{1}
Kaiqi Kou$^{*}$\textsuperscript{1}
Shaowei Wang$^{*}$\textsuperscript{1}
Ruoyu Wang\thanks{Equal contribution, listed in the random order; \textsuperscript{$\dagger$}Corresponding authors: \{lixiaohui33,baihaoli\}@huawei.com.}~~\textsuperscript{2}
Xiaohui Li\textsuperscript{$\dagger$ 1} \\
\textbf{Sidi Yang\textsuperscript{3}
Yiming Du\textsuperscript{4}
Jianbo Dai\textsuperscript{1}
Zhiming Mao\textsuperscript{4}
Xinyu Wang\textsuperscript{1} 
Lifeng Shang\textsuperscript{1}
Haoli Bai\textsuperscript{$\dagger$ 1}} \\
\textsuperscript{1}Huawei Technologies, 
\textsuperscript{2}NTU, 
\textsuperscript{3}HKU,
\textsuperscript{4}CUHK 
\\
\footnotesize
\includegraphics[height=1em]{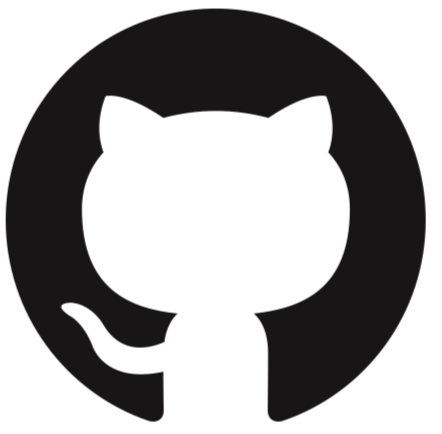}\ \url{https://github.com/SWE-Lego/SWE-Lego}
\quad  
\includegraphics[height=1em]{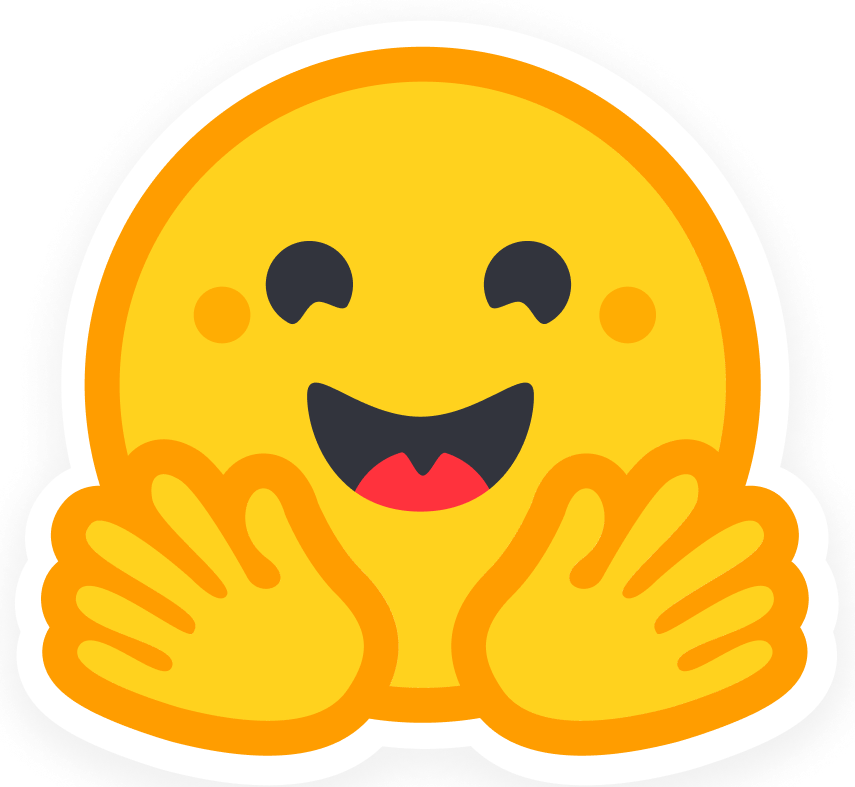}\ \url{https://huggingface.co/SWE-Lego}
}

\renewcommand{\headeright}{Technical Report}
\begin{document}

\maketitle
\thispagestyle{fancy}

\newcommand{\fix}{\marginpar{FIX}}
\newcommand{\new}{\marginpar{NEW}}

\begin{abstract}
We present \textbf{SWE-Lego}, a supervised fine-tuning~(SFT) recipe designed to achieve state-of-the-art performance in software engineering~(SWE) issue resolving. In contrast to prevalent methods that rely on complex training paradigms (e.g., mid-training, SFT, reinforcement learning, and their combinations), we explore how to push the limits of a lightweight SFT-only approach for SWE tasks. SWE-Lego comprises three core building blocks, with key findings summarized as follows: 1) the \textbf{SWE-Lego dataset}, a collection of \textit{32k} high-quality task instances and \textit{18k} validated trajectories, combining real and synthetic data to complement each other in both quality and quantity; 2) a \textbf{refined SFT} procedure with error masking and a difficulty-based curriculum, which demonstrably improves action quality and overall performance. Empirical results show that with these two building bricks alone, \textit{the SFT can push SWE-Lego models to state-of-the-art performance among open-source models of comparable size}  on SWE-bench Verified: \texttt{SWE-Lego-Qwen3-8B} reaches 42.2\%, and \texttt{SWE-Lego-Qwen3-32B} attains 52.6\%.   3) We further evaluate and improve \textbf{test-time scaling~(TTS)} built upon the SFT foundation. Based on a well-trained verifier, SWE-Lego models can be significantly boosted---for example, 42.2\%$\rightarrow$49.6\% and 52.6\%$\rightarrow$58.8\% under TTS@16 for the 8B and 32B models, respectively.

\end{abstract}

\begin{figure*}[h]
    \centering
    \begin{subfigure}[b]{0.7\textwidth}
        \centering
        \includegraphics[width=\textwidth]{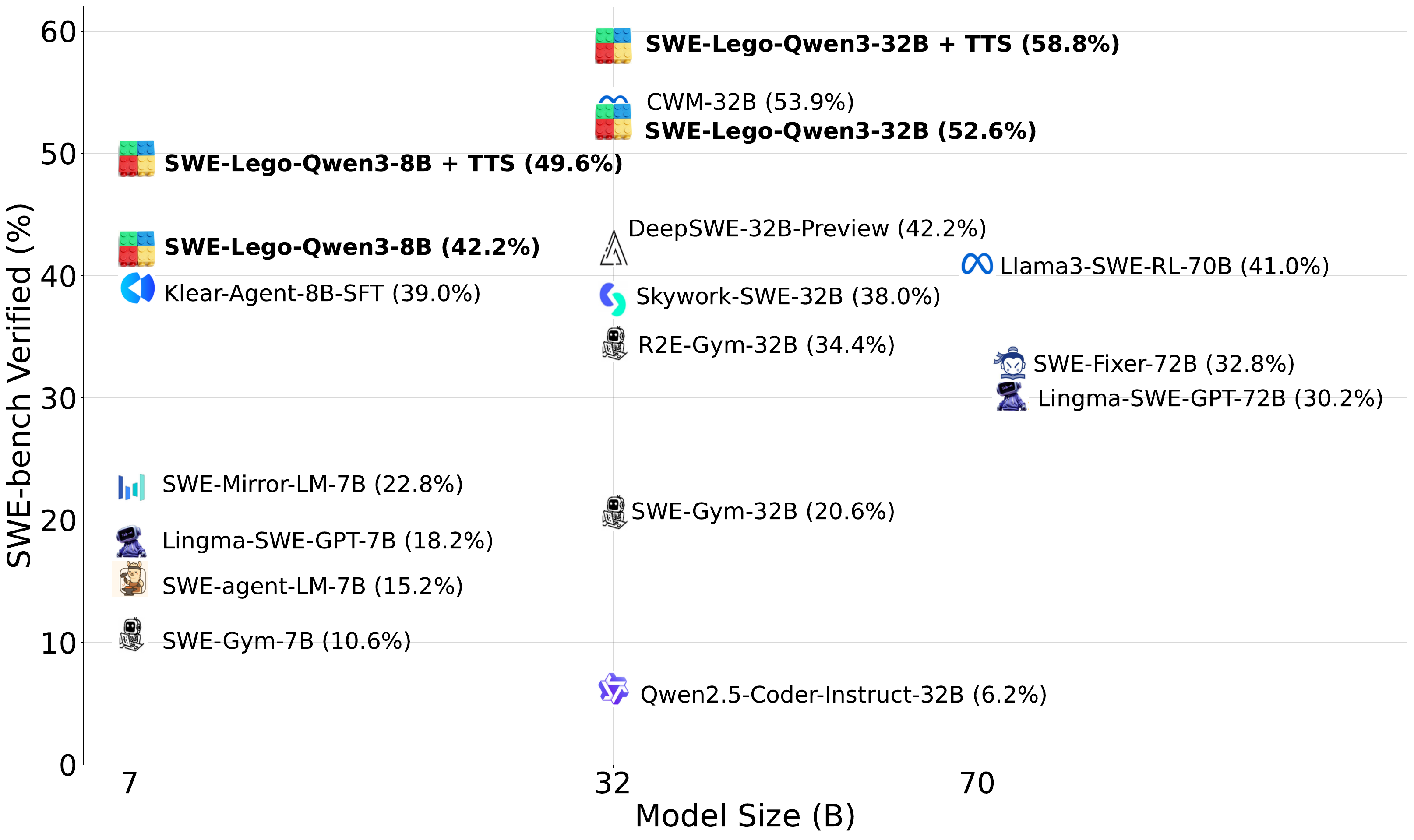}
        \caption{Performance comparisons.}
        \label{fig:perf_comparisons}
    \end{subfigure}
    \hfill
    \begin{subfigure}[b]{0.29\textwidth}
        \centering
        \includegraphics[width=\textwidth]{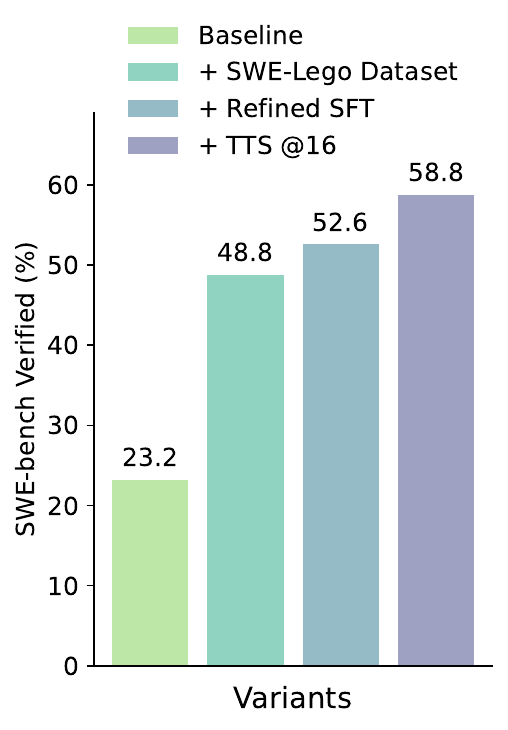}
        \caption{Performance breakdown.}
        \label{fig:incre_gains}
    \end{subfigure}
    \caption{
    Overview of the performance by SWE-Lego models and its breakdown analysis. (a) SWE-Lego models establish a new frontier on SWE-bench Verified, outperforming same-scale competitors. Notably, our results are based on hack-free evaluation, whereas prior works scores could be inflated by the Git hacking. (b) Our hybrid SWE-Lego dataset delivers the largest boost of +25.6\%; subsequent refined SFT adds +3.8\%, and TTS contributes +6.2\%, together lifting the Qwen3-32B model from 23.2\% to 58.8\%.
    }
    \label{fig:into_overall}
\end{figure*}

\section{Introduction}
\label{sec:intro}
Recent advances of large language models (LLMs) have led to remarkable progress in autonomous software engineering (SWE) agents, capable of repository-scale issue resolving~\citep{yang2024sweagent,zhang2024autocoderover,xia2024agentless}. Unlike short, single-file tasks~\citep{du2023classeval,austin2021program,chen2021evaluating,li2022competition}, SWE agents are able to navigate code repositories, faithfully reproduce failing tests, identify buggy files and lines, implement fixes, and validate within executable sandboxes~\citep{yang2023intercode,liu2024repobench,shrivastava2023repofusion}, which demands long-horizon reasoning, robust tool use, and multi-turn interaction abilities.

To equip LLMs with these capabilities, a common approach is supervised fine-tuning (SFT), which distills knowledge from stronger SWE agents. However, prior efforts often lack datasets that include execution environments and realistic bug instances needed to produce high-quality, diverse trajectories~\citep{wang2025swe,yang2025swesmith,jain2025r2egym}. Beyond SFT, performance on SWE tasks can be further improved via mid-training, but this typically requires substantially more computation and data~\citep{copet2025cwm,yang2025kimi}. Alternatively, recent work applies reinforcement learning (RL) without a fixed teacher model or predefined SWE trajectories~\citep{deepswe2025,sonwane2025bugpilot,wei2025toward}, yet limited executable instances remain a bottleneck. In addition, RL is more computationally intensive, demands heavier training infrastructure, and is prone to training collapse due to hyperparameter sensitivity. These practical constraints motivate a critical question: \emph{how far can we push a lightweight, SFT-only training recipe for SWE tasks?}

To investigate, we propose \textbf{SWE-Lego}, a simple yet powerful SFT framework for training SWE agents. We discover that with careful data design, training techniques, and test-time scaling for inference, SFT alone can achieve state-of-the-art performance, matching or surpassing more complex training paradigms under similar LLM scales, as shown in Figure~\ref{fig:into_overall}. SWE-Lego comprises three core building blocks:

\begin{itemize}[leftmargin=2ex]
\setlength{\itemsep}{2pt}
\item We introduce the \textbf{SWE-Lego dataset}, a collection of \textbf{32k} high-quality task instances and \textbf{18k} validated expert trajectories. The SWE-Lego dataset is constructed from two complementary sources: high-quality real-world GitHub pull requests~\citep{badertdinov2025swerebench}, and scalable synthetic instances~\citep{yang2025swesmith}, where both sources complement each other in quality and
quantity: the real-world data provides the majority of benefits but is hard to scale, while the synthetic data provides
additional gains with further scaling. Furthermore, we implement rigorous trajectory validation and filtering, 
such as preventing agents from Git hacking (\textit{i.e.}, inspecting Git history to recover solutions) to ensure the data teaches genuine problem-solving skills.
\item We refine the conventional supervised fine-tuning to better train SWE agents. While standard SFT approaches have been widely applied for training~\citep{yang2025swesmith,wang2025swe},
they typically treat every token in an expert trajectory equally during loss calculation. This overlooks a critical issue: trajectories often contain both successful actions and intermediate errors, and naively learning from mistakes could reinforce undesired behavior. 
To address this, we introduce \textbf{{step-level error masking}}, which excludes tokens associated with execution-time errors (e.g., tool failures or failing tests) from the loss. Moreover, we introduce another refinement of \textbf{{difficulty-based curriculum learning}}, which trains the model on easier tasks before progressing to harder ones, with difficulty estimated empirically by trajectory length.
\item To trade additional compute for further gains, we study \textbf{test-time scaling (TTS)} ~\citep{jain2025r2egym,openhands_critic_model,deepswe2025} along two orthogonal dimensions: (i) sequential scaling by increasing the maximum number of interaction turns, and (ii) parallel scaling via multiple rollouts with verifier selection.
We seek to explore the best TTS strategies for SWE tasks given the compute constraints. We find that sequential scaling saturates around 100–140 turns; beyond this point, compute is better allocated to parallel rollouts. In addition, generative verifiers~\citep{swe-gym,jain2025r2egym} consistently outperform regression-based scorers~\citep{openhands_critic_model} for parallel scaling.
\end{itemize}

Empirically, SWE-Lego establishes new state-of-the-art results among open-source models on SWE-bench Verified~\citep{jimenez2024swebench}: SWE-Lego-Qwen3-8B reaches 42.2\%, and SWE-Lego-Qwen3-32B attains 52.6\% using only SFT (Figure~\ref{fig:perf_comparisons}). With test-time scaling, the performance rises to 49.6\% and 58.8\% under TTS@16, respectively. These results demonstrate that a meticulously crafted SFT pipeline can rival or exceed the performance of more complex and computationally expensive training regimes. Notably, our results are obtained without Git hacking, whereas prior reports are often inflated by such solution leakage.
Ablations in Figure~\ref{fig:incre_gains} show that our SWE-Lego dataset yields a substantial boost of +25.6\%, refined SFT contributes an additional +3.8\%, and TTS ultimately adds +6.2\%, which together lift Qwen3-32B from 23.2\% to 58.8\%, confirming the effectiveness of each building blocks.
We hope SWE-Lego can provide a reproducible and lightweight post-training paradigm for building effective SWE agents, and the associated data and models of this work will be open-sourced for future research in the community.

\section{\textsc{SWE-Lego} Dataset: Combining Real-world and Synthetic Data}

\subsection{Overview}

LLMs are increasingly integrated into autonomous agents for real-world SWE tasks. We adopt the SWE-Agent paradigm~\citep{yang2024sweagent}, in which an LLM operates through a specialized agent-computer interface that allows it to autonomously navigate repositories, search and edit code, and execute tests to diagnose and resolve issues. In this context, a \emph{SWE task instance} is usually defined as the concrete SWE issue within a codebase, comprising a natural-language problem description, test cases, a reference golden patch, and related artifacts.
Besides, a \emph{trajectory} is the complete sequence of actions taken by an agent and the interleaved observations from environments. 
To train capable SWE agents, \emph{our primary goal is to curate a large-scale, diverse, and executable collection of task instances, together with high-quality, validated trajectories}.

\begin{table}[t]
  \centering
  \resizebox{1.0\linewidth}{!}{
  \renewcommand{\arraystretch}{1.4}
  \begin{tabular}{lccccc}
    \toprule
    \textbf{Dataset} & \textbf{Type} & \textbf{Executable} & \textbf{Repositories} & \textbf{Task Instances} & \textbf{Valid Trajectories} \\
    \midrule
    R2E-Gym-Train & Synthetic & Yes & 10 & 4.6k & 3.3k \\
    SWE-Swiss & Synthetic & Yes & $<$600  & 10k & NA \\
    SWE-Fixer & Real & No & 856 & 115k & NA \\
    SWE-Gym & Real & Yes & 11 & 2.4k & 491 \\
    SWE-bench-Train & Real & No & 37 & 19k & NA \\ 
    SWE-rebench & Real & Yes & 3.5k & 21k & NA \\
    SWE-smith & Real \& Synthetic & Yes & 128 & 50k & 5k \\
    \cmidrule(lr){1-6}
    \textbf{SWE-Lego} & Real \& Synthetic & Yes & 3251 & 32.1k & 14.1k resolved + 4.0k semi-resolved \\
    \bottomrule
  \end{tabular}
  }
  \caption{Comparison of public SWE issue-resolving datasets and our SWE-Lego dataset. The proposed dataset combines real and synthetic instances at scale, with executable environments and a large pool of validated trajectories, providing a stronger foundation for training SWE agents. Note that SWE-Swiss and SWE-Fixer release only sub-trajectories, and therefore, counts of full trajectories are not reported. ``Semi-resolved'' means the trajectories do not pass all test cases but have partial correctness, \textit{e.g.}, correct file localization.}
  \label{tab:dataset_comparison}
\end{table}

\begin{figure}[t]
  \centering
  \includegraphics[width=\linewidth]{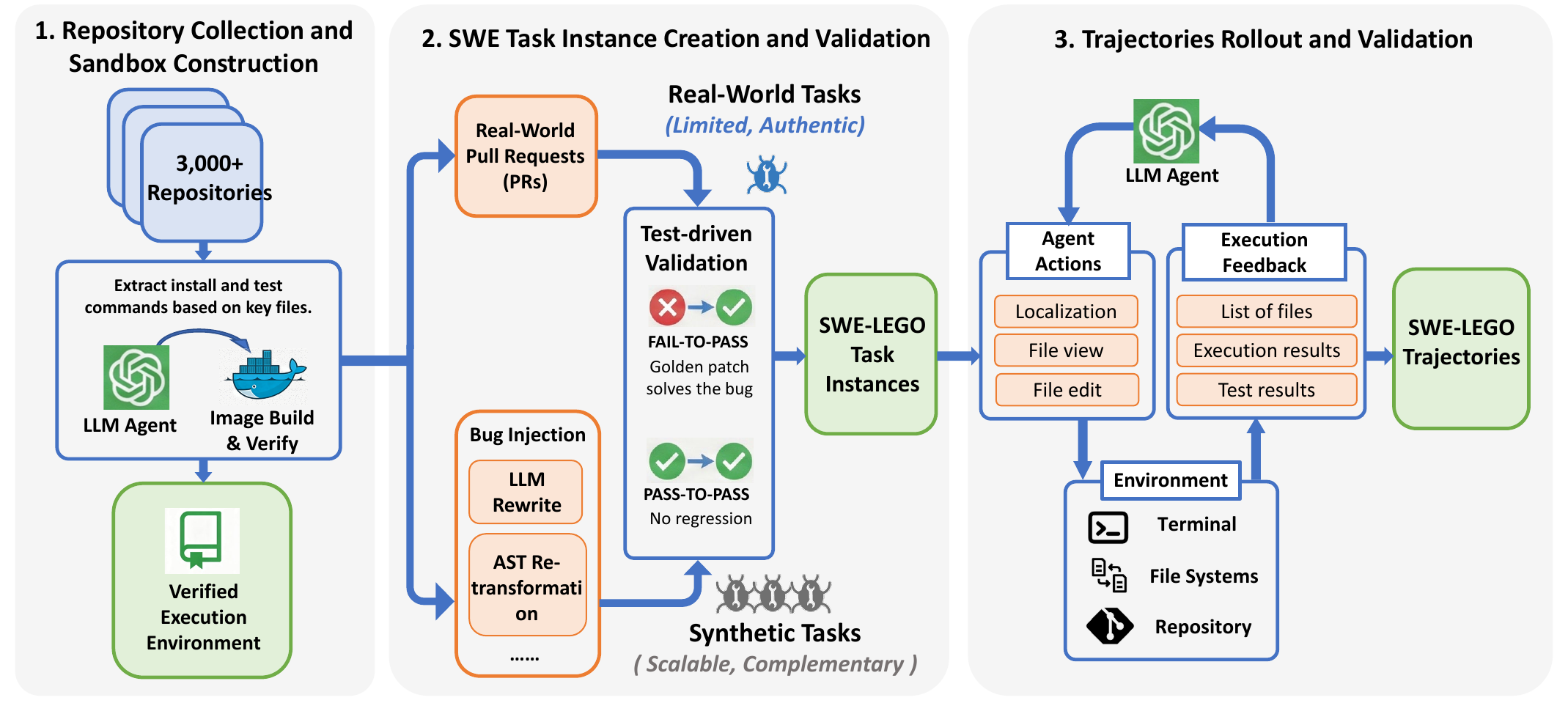} 
  \caption{
  Our SWE‑Lego pipeline comprises three stages: environment construction from over 3,000 repositories; hybrid task creation by combining real pull requests with synthetic bugs; and expert‑trajectory generation and curation for SFT.
  }
  \label{fig:swe_lego_pipeline}
\end{figure}

\begin{figure*}[t]
    \centering
    \begin{subfigure}[b]{0.49\textwidth}
        \centering
        \includegraphics[width=\textwidth]{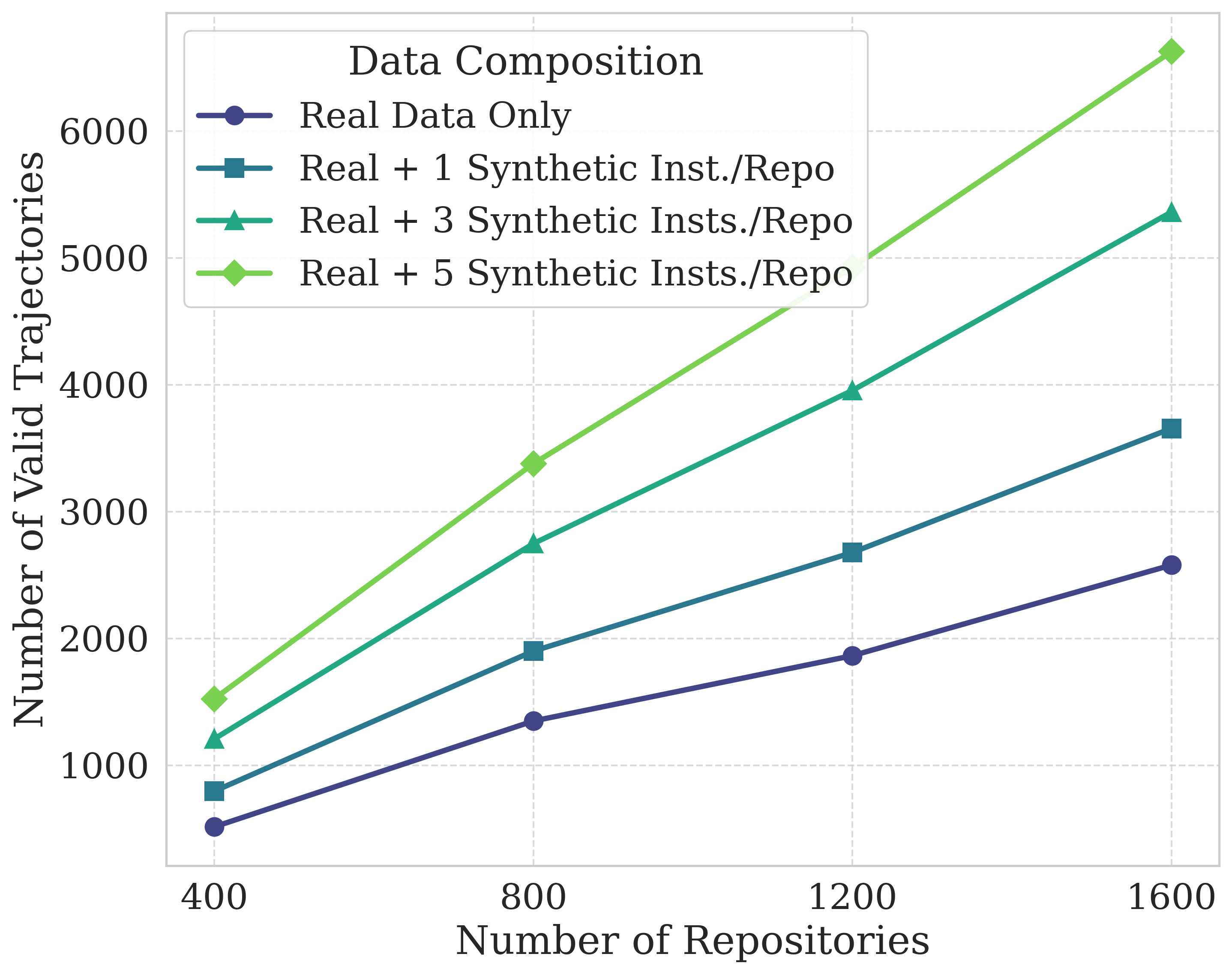}
        \caption{Scaling of Valid Trajectories}
        \label{fig:traj_scaling}
    \end{subfigure}
    \hfill
    \begin{subfigure}[b]{0.49\textwidth}
        \centering
        \includegraphics[width=\textwidth]{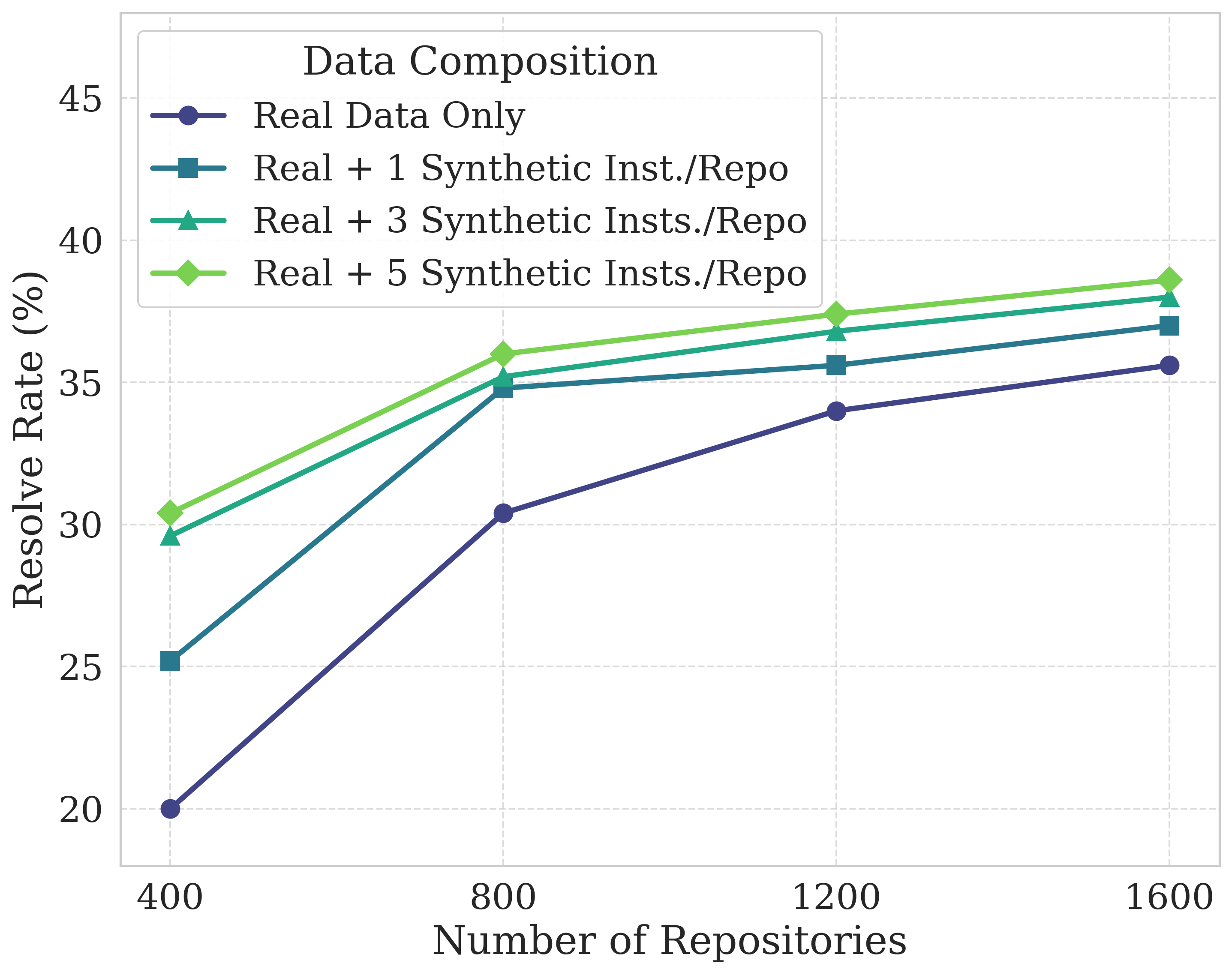}
        \caption{Performance Scaling on SWE-bench Verified}
        \label{fig:perf_scaling}
    \end{subfigure}
    \caption{Impact of synthetic data augmentation on data scale and performance. (a) The number of valid expert trajectories increases substantially with the addition of synthetic instances, demonstrating an effective way to expand supervision beyond limited real-world PRs. (b) The resolve rate on SWE-bench Verified rises as more synthetic instances are added per repository, across different repository counts, indicating that hybrid data improves not only dataset scale but also downstream model effectiveness. ``Insts./Repo'' denotes the average number of instances per repository. 
    }
    \label{fig:scaling_analysis}
\end{figure*}

We adopts a hybrid data construction strategy that combines real-world and synthetic SWE task instances, as shown in Figure~\ref{fig:swe_lego_pipeline}.  This is motivated by the empirical observation that neither data source alone is sufficient: real-world instances, while authentic, are inherently limited in quantity given strict filtering criteria. On the other hand, synthetic instances, while scalable, lack the complexity of natural software repositories (Table~\ref{tab:task_instance_statistics}). 
We therefore mix the real with synthetic instances, and apply rigorous generation and validation procedures to produce high-quality trainable trajectories.

The resulting SWE-Lego dataset comprises 32k executable task instances and 18k validated expert trajectories to date, as summarized in Table~\ref{tab:dataset_comparison}. Compared with existing open-source SWE datasets, SWE-Lego spans more repositories, offers more executable task instances, and contains substantially more validated trajectories. Moreover, scaling this hybrid corpus consistently yields better downstream performance on SWE-bench Verified, as evidenced in Figure~\ref{fig:scaling_analysis}. In the following subsections, we detail the pipeline used to construct the SWE-Lego dataset.

\subsection{SWE Task Instance Creation and Validation}
\label{sec:swe-task}

\paragraph{Repository Collection and Sandbox Construction.}
The proposed dataset is built upon SWE-rebench~\citep{badertdinov2025swerebench}, a large collection of real-world executable repositories. We select over 3,000 Python-centric repositories with permissive licenses and active maintenance histories. To ensure reproducibility, we deploy a fully automated pipeline that parses configuration files (e.g., \texttt{setup.py}) to build Docker containers. Only repositories that successfully build and pass sanity tests are retained, serving as the verifiable ``base plates'' for our subsequent task construction.

\paragraph{Real-world and Synthetic Task Construction.}
\emph{Real-world tasks} are derived from resolved GitHub pull requests (PRs) paired with corresponding issue descriptions. They offer high authenticity with production-level bug complexity, but are often labor-intensive and limited in quantity due to the strict filtering criteria~\citep{badertdinov2025swerebench,swe-gym,jain2025r2egym,zeng2025skywork}. Moreover, each real-world task requires a unique sandbox aligned with the pre-fix snapshot. While SWE-smith~\citep{yang2025swesmith} proposes a ``PR mirror'' method to generate real-world instances (prompting LLMs to revert PR edits via .diff plaintext instead of checking out the base commits), we find this approach yields lower-quality data compared to directly collecting PRs that satisfy SWE-rebench~\citep{badertdinov2025swerebench}’s filtering rules. Thus, we adopt the latter strategy for real-world task construction.

\emph{Synthetic tasks}, by contrast, are generated via active bug injection following SWE-smith~\citep{yang2025swesmith}, leveraging two core techniques: (i) \emph{LLM Rewrite}: prompting models to rewrite code using only function headers and docstrings; (ii) \emph{AST Reformulation}: extracting abstract syntax trees (ASTs) for classes/functions and applying random transformations, \textit{e.g.}, removing conditionals/loops, modifying operators or dependencies. Synthetic tasks enable high scalability and efficiency, as they are not constrained by the availability of filtered real PRs. Even repositories with no qualified PRs can be used to create large-scale tasks at low cost. Additionally, synthetic tasks from the same repository share a single sandbox environment, since all bugs are injected into the same codebase version.

These two data sources are complementary for scaling SWE tasks: real-world data provides \emph{depth} (complexity and realism), while synthetic data provides \emph{breadth} (quantity and coverage). Table~\ref{tab:core_fields_derivation} summarizes the core-field differences between the two, and additional implementation details are provided in Appendix~\ref{sec:apdx-real-syth-data}.
Table~\ref{tab:task_instance_statistics} presents a quantitative comparison of complexity. We observe a stark contrast: real-world instances touch more files and change more lines than synthetic ones. This confirms that real-world data captures the sprawl of complex bugs, whereas synthetic data offers focused, surgical training signals and yields a higher rate of valid trajectories for training. Further analysis of issue categories is provided in Appendix~\ref{sec:issue-category-analysis}. Figure~\ref{fig:scaling_analysis} demonstrates the effectiveness of this hybrid approach: for a fixed set of repositories, scaling synthetic data, \textit{e.g.}, from ``real only'' to ``real + 5 synthetic insts./repo'', consistently improves both the number of valid expert trajectories and the resolve rate of the trained models. Together with Table~\ref{tab:dataset_comparison}, this shows that scaling executable hybrid data, rather than real instance alone, is key to obtaining richer supervision signals and higher downstream performance.

\begin{table}[t]
  \centering
  \small
    \begin{tabular}{@{}l|p{4.5cm}|p{4.5cm}@{}}
      \toprule
      \textbf{Core Field} & \textbf{Real-World Tasks} & \textbf{Synthetic Tasks} \\
      \midrule
      \texttt{golden\_patch} & Merged PR diff as the ground-truth fix. & Inverse the patch of  the programmatically injected bug. \\
      \midrule
      \texttt{FAIL-TO-PASS} \& \texttt{PASS-TO-PASS} & Label of test sets that induced by running the original project tests before and after the merged patch. & Label of test sets that  induced by running the same tests on buggy vs. restored (pre-injection) code.\\
      \midrule
      \texttt{problem\_statement} & Original GitHub issue linked to the PR. & LLM-generated description conditioned on the golden patch, test files, and logs. \\
      \midrule
      \texttt{image\_name} (Sandbox)& Per-PR sandbox tied to the exact pre-merge commit. & Shared sandbox for many injected bugs on the same base commit given the repository.\\
      \bottomrule
    \end{tabular}
  \caption{High-level comparison of how core fields are instantiated for real-world and synthetic SWE tasks in SWE-Lego. Both sources populate the same unified schema, but differ in data origin and sandbox granularity, leading to realistic yet costly real instances and scalable, controllable synthetic ones.}
  \label{tab:core_fields_derivation}
\end{table}

\begin{table}[t]
  \centering
  \small
  \renewcommand{\arraystretch}{1.}
  \setlength{\tabcolsep}{4.pt}
  \begin{tabular}{lcccc}
    \toprule
    \textbf{Category} & \textbf{Metric} & \textbf{Real-World Tasks} & \textbf{Synthetic Tasks} & \textbf{Real-World + Synthetic Tasks} \\
    \midrule
    \textbf{Instances} & -- & 18409 & 13710 & 32119 \\
    \cmidrule(lr){1-5}
    \textbf{Issue Length} & Words & 135 & 168 & 149 \\
    \cmidrule(lr){1-5}
    \multirow{3}{*}{\textbf{Golden Patch}} & Files & 3.7 & 1.0 & 2.6 \\
    & Hunks & 9.5 & 1.3 & 6.0 \\
    & Lines & 138.0 & 18.8 & 87.3 \\
    \cmidrule(lr){1-5}
    \multirow{3}{*}{\textbf{Test Cases}} & Fail-to-Pass & 12.4 & 10.9 & 11.7 \\
    & Pass-to-Pass & 74.6 & 41.2 & 60.4 \\
    & Total & 87.0 & 52.1 & 72.1 \\
    \midrule
    \textbf{Valid Trajectories} & -- & 5010&9100 &14110 \\
    \bottomrule
  \end{tabular}
  \caption{Characteristic comparisons of task instances. Real-world tasks are more complex, touching more files and modifying more lines, while synthetic tasks are more focused and yield more valid, fully resolving trajectories. The combined dataset covers a broader spectrum of difficulty.
  }
  \label{tab:task_instance_statistics}
\end{table}

\subsection{Trajectory Rollout and Validation}
\label{sec:high_quality_traj}

\begin{figure}[t]
  \centering
  \includegraphics[width=.99\linewidth]{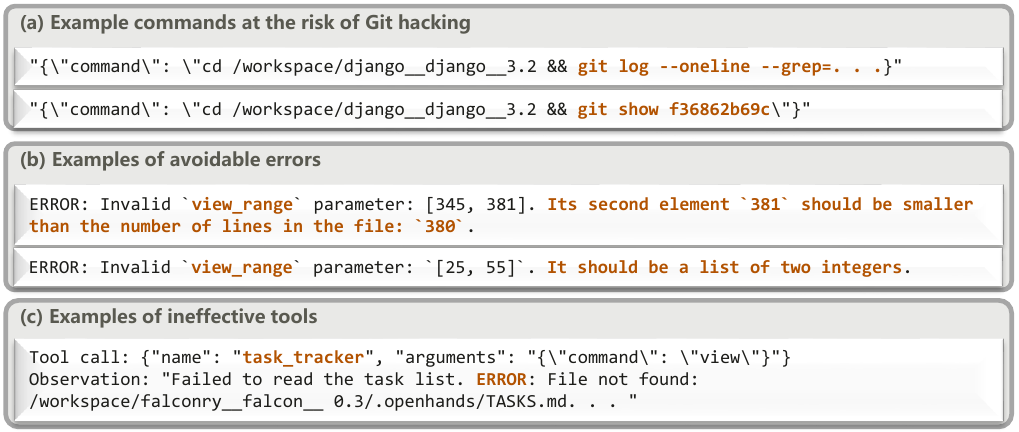}
  \caption{
  Examples of problematic commands or tool interactions: (a) high‑risk commands that can cause Git hacking; (b) \texttt{view\_range} parameter mis-specification; (c) ineffective \texttt{task\_tracker}.}
  \label{fig:traj_generation_practices}
\end{figure}

With the hybrid task instances, we now roll out expert trajectories that are later used for SFT. 
We adopt the \texttt{OpenHands} scaffold (v0.53)~\citep{wang2024openhands} with \texttt{Qwen3-Coder-480B-A35B-Instruct}~\citep{qwen3} as the teacher agent, and set the maximum number of interaction turns to 100. The model offers competitive coding and reasoning capabilities on challenging SWE tasks and, being open-weight, can be deployed locally at scale for trajectory distillation. During distillation, we found that the agent sometimes peeked at Git history or future commits, thereby cheating rather than faithfully resolving the issue. As showcased in Figure~\ref{fig:traj_generation_practices}, we also observed avoidable tool-calling errors and tools that were ineffective in practice. To enhance trajectory quality, we propose and implement the following  three practices.

\textbf{Preventing Git Hacking.} Recently it is known by the SWE community that 
LLM agents can unexpectedly ``hack'' git metadata to locate the golden patch by directly inspecting commit logs\footnote{\url{https://github.com/SWE-bench/SWE-bench/issues/465}, posted on Sep 3, 2025. }.
A demonstrative example is shown Figure~\ref{fig:traj_generation_practices}(a), and its associated full trajectory is listed in Appendix~\ref{sec:git_leak_full_example}.
To prevent such Git hacking, we sanitize the repository history exposed to the agent. For real instances, we remove all commits and log messages dated after the issue creation date so that future fixes are invisible. For synthetic instances, since the buggy version of repository is ahead from the non-buggy version due to intentional bug injection, we remove the entire git history and all logs, exposing only a single snapshot of the buggy codebase. This forces the agent to genuinely reason about code and tests rather than reading answers from version control. As expected, we see in Table~\ref{tab:practice_traj_generation} that the valid-trajectory rate drops slightly from an inflated 31\% to 30\%, and the average number of turns increases a bit from 61.8 to 62.5. In addition, as mentioned in Section~\ref{sec:main_results}, we also report the results of SWE-bench Verified without Git hacking throughout this paper, unless specified otherwise. 

\textbf{Handling Malformed Tool Errors.}
With \texttt{Qwen3-Coder-480B-A35B-Instruct}, we observe frequent malformed calls to the \texttt{str\_replace\_editor} tool, \textit{e.g.}, passing a string to \texttt{view\_range} or specifying out-of-bounds line ranges (see Figure~\ref{fig:traj_generation_practices}(b)), causing tool failures and a waste of interaction budget. To mitigate these errors, we apply lightweight post-hoc correction, \textit{i.e.}, parsing strings to integers and clipping ranges to valid spans, which enables the agent to inspect code more reliably. Without robust tool handling, the valid-trajectory rate decreases slightly from 30\% to 29\%, and the average number of turns increases from 62.5 to 66.6 (see Table~\ref{tab:practice_traj_generation}).

\textbf{Pruning Ineffective Tools.}
Recent work~\citep{openai_gpt51_codex_max,anthropic_claude_sonnet_4_5} has begun integrating task-management tools such as \texttt{task\_tracker} for long-horizon planning and progress tracking. However, we find that \texttt{Qwen3-Coder-480B-A35B-Instruct} struggles to use them effectively, which often cause execution errors (see Figure~\ref{fig:traj_generation_practices}(c)). We therefore discard this tool and restrict the tool set to four essential operations: \texttt{execute\_bash}, \texttt{str\_replace\_editor}, \texttt{think}, and \texttt{finish}, to keep trajectories streamlined, as evidenced by fewer interaction turns (from 65.5 to 62.5) and a higher valid-trajectory rate (from 27\% to 30\%) in Table~\ref{tab:practice_traj_generation}.

\paragraph{Validation and Filtering.}
Quality control is enforced through rigorous post-hoc validation. We classify trajectories as \emph{resolved} if they pass all tests without regression. However, not all resolved trajectories are equal, nor are all unresolved ones useless.
As shown in Table~\ref{tab:data_filering_resolved_rates}, our filtering strategy is two-fold:
(i) we filter low-quality resolved trajectories (\textit{e.g.}, those that ``cheat'' by modifying tests), which yields a modest performance gain;
(ii) we recycle semi-resolved trajectories, \textit{i.e.}, those that correctly locate the buggy file (with the recall rate of 100\%) but fail to fix it. It results 4k additional trajectories for training. Adding these trajectories provides a clear boost (+1.2\%), demonstrating that even partial successes contain valuable supervision for fault localization.

\begin{table}[t]
    \centering
    \begin{minipage}{0.55\textwidth}
        \centering
        \resizebox{\linewidth}{!}{
        \begin{tabular}{lcc}
        \toprule
        \textbf{Practice} & \textbf{Avg Turns} & \textbf{Valid Traj. Rate (\%)} \\
        \midrule
        All  & 62.5  & 30  \\
        w/o preventing Git hacking & 61.8  & 31 \\
        w/o handling malformed tool errors & 66.6 & 29 \\
        w/o pruning ineffective tools  & 65.5 & 27  \\
        \bottomrule
        \end{tabular}
        }
        \caption{Ablation of rollout practices. The valid trajectory rate is evaluated across 100 SWE-Lego instances. 
        }
        \label{tab:practice_traj_generation}
    \end{minipage}
    \hfill
    \begin{minipage}{0.42\textwidth}
        \centering
        \resizebox{\linewidth}{!}{
        \begin{tabular}{lcc}
        \toprule
        \textbf{Data Strategy} &  \textbf{\# Traj} &  \textbf{Resolve Rate (\%)} \\
        \midrule
        All resolved   &  14.6k  & 40.4 \\
        - low-quality resolved   &  14.1k  & 41.0 \\
        + semi-resolved      & 18.1k &  42.2 \\ 
        \bottomrule
        \end{tabular}
        }
        \caption{Ablation of filtering strategies on SWE-bench Verified.}
        \label{tab:data_filering_resolved_rates}
    \end{minipage}
\end{table}

\begin{mybox}[colback=gray!10]{Takeaways}
    \begin{itemize}[leftmargin=2ex]
      \item We introduce the \textbf{SWE-Lego} dataset, comprising  \textbf{32k} high-quality software engineering task instances and \textbf{18k} expert trajectories, the core building brick for training effective SWE agents. \\
      \item We demonstrate that combining real-world and synthetic data yields complementary benefits in both \textbf{quantity and quality}. Moverover, trajectory curation and filtering further enhance data utility, thereby improving the issue resolving rate.
    \end{itemize}
\end{mybox}

\section{Refined Supervised Fine-tuning}

Conventional supervised fine-tuning on expert trajectories has proven effective for SWE tasks, however, the intrinsic characteristics of these trajectories remains underexplored. A trajectory may ultimately resolve an issue, yet it often contains erroneous intermediate steps, \textit{e.g.}, failed attempts and malformed tool calls. Rather than blindly learning from the entire trajectory, we propose \emph{step-level error masking}, which enables the model to learn from correct actions while excluding incorrect ones. Furthermore, we investigate \emph{curriculum learning} in the SWE context, wherein progressively increasing trajectory difficulty better guides the training process.

\subsection{Step-level Error Masking}
\begin{figure}[htbp]
  \centering
  \includegraphics[width=\linewidth]{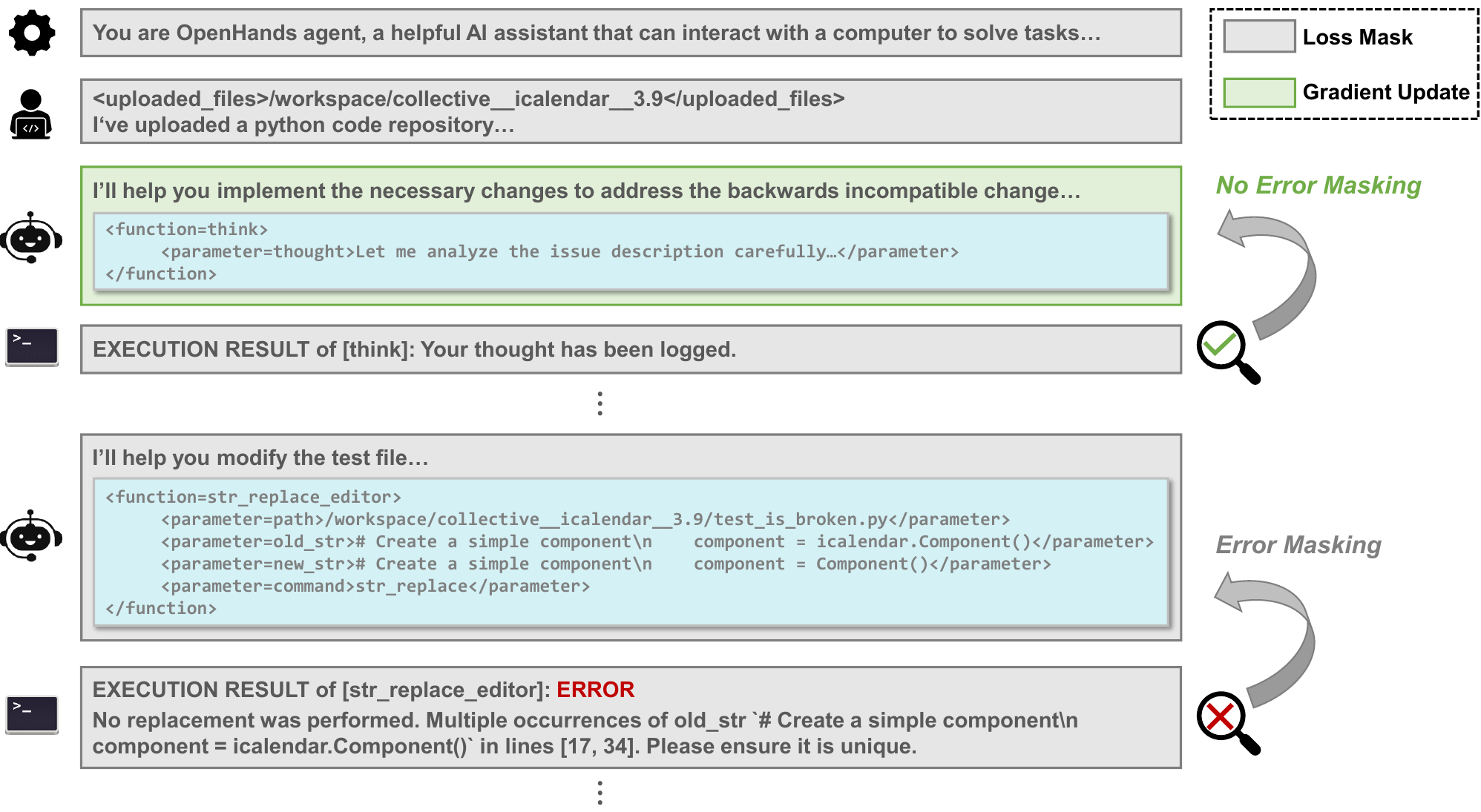}
  \caption{An example of step-level error masking, which maintains the complete trajectory context while selectively masking the loss calculation on incorrect agent responses.}
  \label{fig:error_mask}
\end{figure}

While the high-quality trajectories identified in Section \ref{sec:high_quality_traj} offer valuable learning signals at a macro level, we propose a more granular refinement: step-level error masking. This approach is motivated by the observation that expert demonstrations rarely proceed monotonically to success; they often contain intermediate missteps, such as invalid function calls or incorrect arguments, that are later identified and corrected. Our method {maintains the full trajectory context but selectively masks the loss calculation on erroneous agent responses}.
As illustrated in Figure \ref{fig:error_mask}, we use regular expression to identify error messages provided by the terminal environment, and apply error masking to the corresponding agent response.
Crucially, errors stemming from reproducing bugs or executing test files are excluded from this masking process. This technique applies gradient updates solely to valid actions, thereby providing a robust learning signal that enhances both action generation and error recovery capabilities.
By emphasizing learning from correct actions, it directly reduces core reasoning failures, such as ``Incorrect Implementation'' and ``Localization Error'', that are prevalent in later training stages (Section \ref{sec:error_analysis}). 
Empirical results in Table \ref{tab:training_strategy_resolved_rates} demonstrate that step-level error masking improves model performance by over 2 points.

\subsection{Difficulty-Based Curriculum Learning}
\begin{figure}[t]
  \centering
  \begin{minipage}[!t]{0.45\textwidth} 
    \centering
    \includegraphics[width=\linewidth]{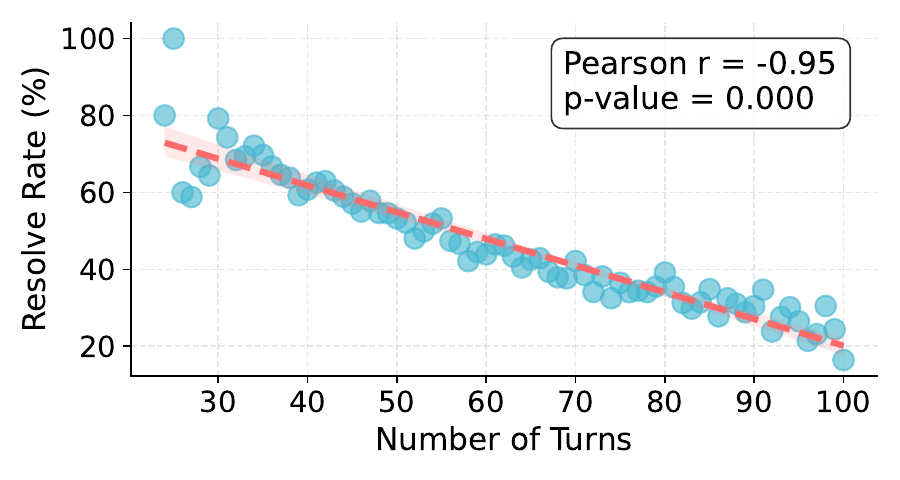} 
    \vspace{-6mm}
    \caption{Correlation between number of turns and average resolve rate.} 
    \label{fig:corr_turn_resolved}
  \end{minipage}
  \hfill 
  \begin{minipage}[!t]{0.53\textwidth}
    \centering
    \small
    \resizebox{\linewidth}{!}{
    \renewcommand{\arraystretch}{1.5}
    \begin{tabular}{cc|cc}
  \toprule
  \multirow{2}{*}{\makecell{\textbf{Error} \\ \textbf{Masking}}} & \multirow{2}{*}{\makecell{\textbf{Curriculum} \\ \textbf{Learning}}} & \multicolumn{2}{c}{\textbf{Resolve Rate (\%)}}  \\
        \cmidrule(lr){3-4}
         & & \textbf{SWE-Lego-8B} & \textbf{SWE-Lego-32B} \\
  \midrule
  \ding{51}  & \ding{51} & \textbf{42.2} & \textbf{52.6} \\
  \hline
  \ding{51}  & \ding{55} & 41.8 (\textcolor{academicred}{-0.4}) & 51.8 (\textcolor{academicred}{-0.8}) \\
  \ding{55}  & \ding{51} & 40.2 (\textcolor{academicred}{-2.0}) & 50.4 (\textcolor{academicred}{-2.2}) \\
  \ding{55}  & \ding{55} & 39.4 (\textcolor{academicred}{-2.8}) & 48.8 (\textcolor{academicred}{-3.8}) \\
  \bottomrule
\end{tabular}
}
\captionsetup{type=table}
    \caption{Ablation of training strategies.}
    \label{tab:training_strategy_resolved_rates}
  \end{minipage}
\end{figure}

To maximize sample efficiency and training performance,
we implement a curriculum learning strategy that progressively exposes the model to tasks of increasing complexity. This approach encourages the model to acquire fundamental skills from simpler examples before advancing to more challenging problems, thereby promoting more robust generalization.
We explore two principled methods for difficulty categorization:
\begin{itemize}[leftmargin=2ex]
\item \textbf{Model-Based Scoring:} We fine-tune a Qwen2.5-Coder-32B-Instruct~\citep{hui2024qwen2} model as a difficulty scorer on a dataset of 1.5k human-annotated (task, difficulty) pairs. This scorer achieves 70\% accuracy on a held-out test set and classifies tasks into three tiers: Easy (estimated solving time $<$ 15 minutes), Medium (15 minutes to 1 hour), and Hard ($>$ 1 hour).
\item \textbf{Trajectory-Length-Based Heuristic:} We discover a \emph{strong negative correlation} ($r=-0.95$) between instance resolve rate and the number of turns in the trajectory, as depicted in Figure \ref{fig:corr_turn_resolved}. 
Leveraging this finding, we partition the data into three difficulty bins based on turn count: Easy (0-50 turns), Medium (50-70 turns), and Hard (70-100 turns).
\end{itemize}
For curriculum learning, we select the trajectory-length-based heuristic because (i) it outperformed the model-based approach by 0.5 points, and (ii) it is more computationally efficient and less prone to mislabeling.

We adopt a three-stage SFT curriculum based on difficulty rankings. In each stage, the model is trained exclusively on data from the corresponding difficulty tier. To mitigate catastrophic forgetting of previously acquired skills, the training data for each subsequent stage is augmented with all data from the preceding stages.
This curriculum aligns with the training dynamics observed in Section \ref{sec:error_analysis}: it first grounds the model on ``Easy" tasks to overcome basic ``Failed to Reproduce" errors, then introduces ``Hard" tasks to develop the strategic planning needed to avoid ``Ran Out of Max Turns" failures.
Table \ref{tab:training_strategy_resolved_rates} shows that curriculum learning consistently enhances the performance of both the 8B and 32B models, with the highest results achieved when combined with error masking.

\subsection{Training Details}

For training, we use the \texttt{Qwen3-8B/32B}\footnote{\url{https://huggingface.co/Qwen/Qwen3-8B};\ \ \url{https://huggingface.co/Qwen/Qwen3-32B}}~\citep{qwen3} series as the base models.
We perform full-parameter SFT on the collected SWE-Lego trajectories using LLaMA-Factory framework~\citep{zheng2024llamafactory}.
Models are trained for 4 epochs with a global batch size of 64.
We use the AdamW optimizer~\citep{loshchilov2019decoupledweightdecayregularization} with a weight decay of 0.01 and a cosine learning rate schedule with a
warmup ratio of 0.1.
The learning rate is set to 1e-4 for the 7B/8B models and 5e-5 for 32B models.
To enable training with long-horizon trajectories, we use a maximum context length of 128k tokens.
For the Qwen3 series, RoPE scaling techniques such as YaRN~\citep{peng2023yarnefficientcontextwindow} are adopted to support context lengths of up to 128k tokens.

\subsection{Error Analysis during Training}
\label{sec:error_analysis}

To validate our methodology, we analyze the agent's failure modes during training by categorizing all failed instances (including unresolved, error, and empty patch cases) into five error types based on the agent's trajectory:

\begin{itemize}[leftmargin=2ex]
\setlength{\itemsep}{2pt}
    \item \textbf{Failed to Reproduce}: 
    The agent either did not attempt reproduction or attempted but failed to reproduce the issue.
    \item \textbf{Read Localization Error}: 
    The agent reproduces the issue but fails to open or inspect all files modified in the ground-truth (golden) patch.
    \item \textbf{Write Localization Error}: 
    The agent opens all required files but fails to identify the correct lines to edit, either no edits are made to ground-truth files or the edited lines have no overlap with the ground-truth patch.
    \item \textbf{Ran Out of Max Turns}: The agent successfully reproduced the issue and correctly localized the read and write locations, but reached the maximum iteration limit (100 turns) before completing the task.
    \item \textbf{Incorrect Implementation}: 
    The case does not fall into any of the above categories; the agent appears to find the right location but implements an incorrect fix, resulting in failing tests.
\end{itemize}

\begin{figure}[t]
    \centering
    \begin{subfigure}[b]{0.49\textwidth}
        \centering
        \includegraphics[width=\textwidth]{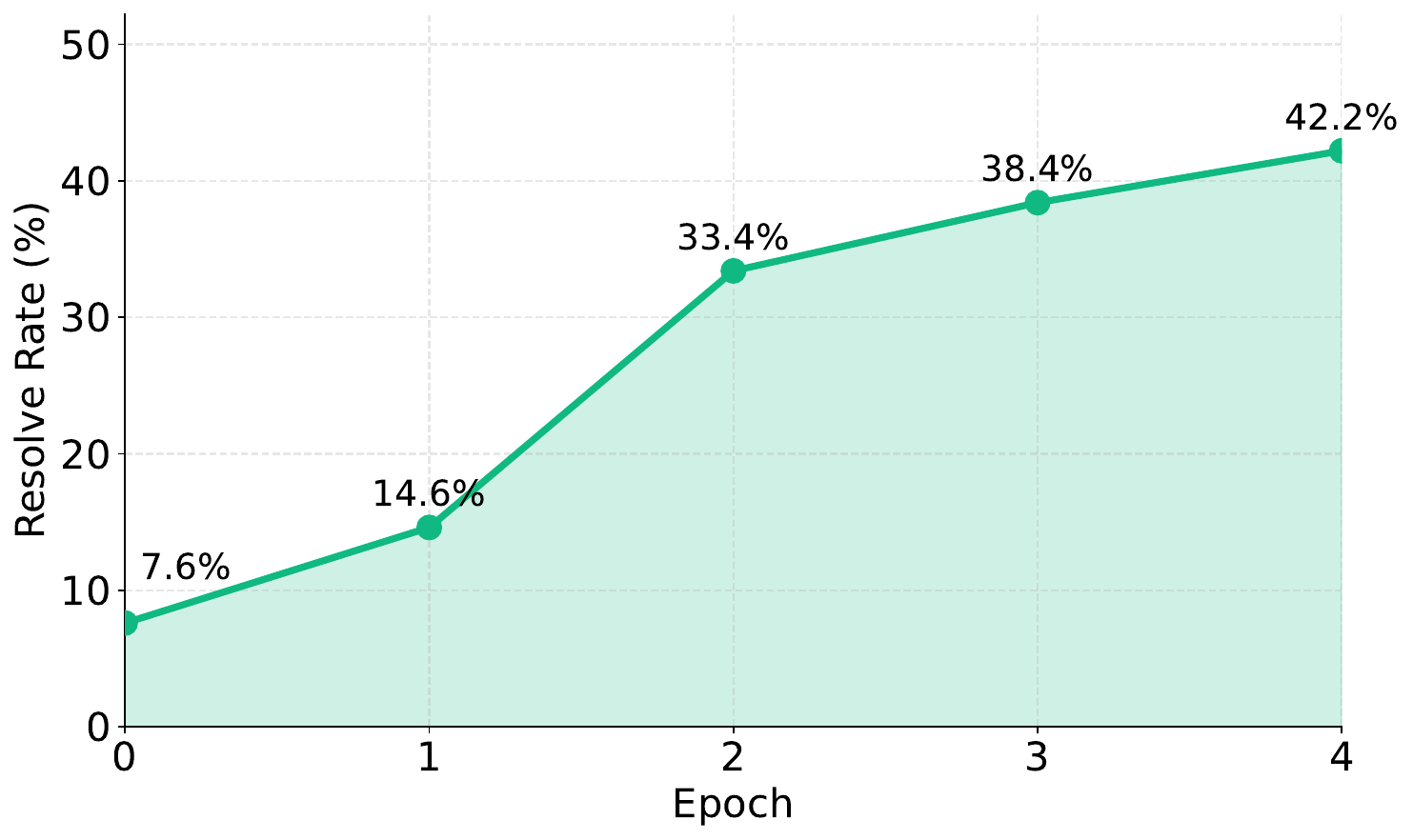}
        \caption{Resolve Rate Improvement Trend}
        \label{fig:resolved_rate}
    \end{subfigure}
    \hfill
    \begin{subfigure}[b]{0.49\textwidth}
        \centering
        \includegraphics[width=\textwidth]{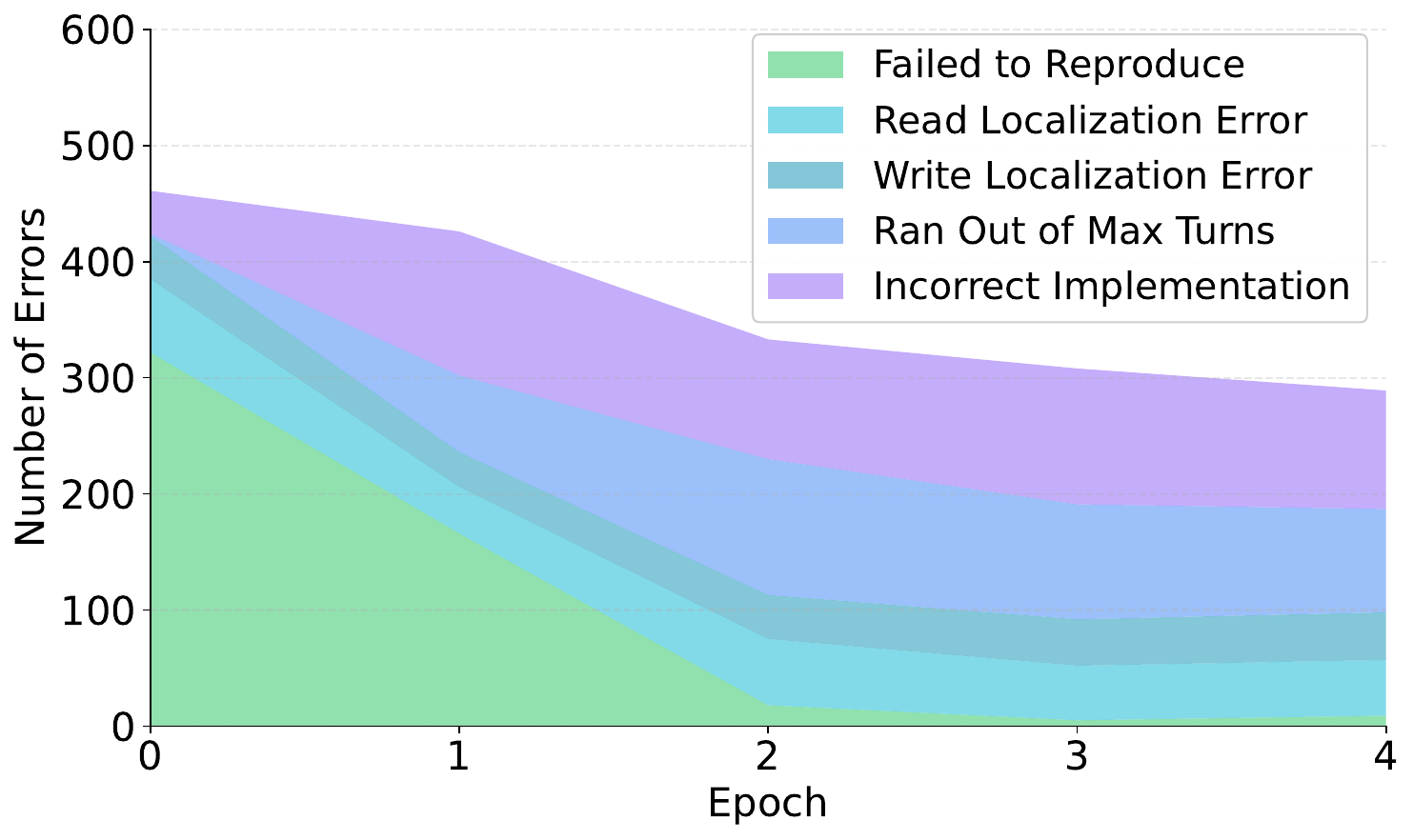}
        \caption{Error Type Distribution Across Epochs}
        \label{fig:error_dist}
    \end{subfigure}
    \caption{Evolution of model performance and failure modes over 4 epochs. (a) Resolve rate increases steadily, reaching 42.2\% by epoch 4.
    (b) Error modes shift from early ``Failed to Reproduce" to mid ``Ran Out of Max Turns" to late  ``Incorrect Implementation" and ``Localization Error" bottlenecks.
    }
    \label{fig:epoch_progression}
\end{figure}

Figure \ref{fig:epoch_progression} shows a clear progression. Early on, ``Failed to Reproduce" errors (38.97\%) dominate, indicating basic task misalignment. After epoch 1, this error vanishes, but ``Ran Out of Max Turns" errors spike (35.14\%), revealing a strategic planning deficit. In later epochs, ``Incorrect Implementation" and ``Localization Error" become the bottlenecks. This shift from process failures to reasoning failures motivates our combined approach: {curriculum learning} sequentially addresses the early and mid-stage errors by structuring task complexity, while {step-level error masking} specifically hones the fine-grained reasoning needed to overcome the persistent late-stage errors.

\begin{mybox}[colback=gray!10]{Takeaways}
    \begin{itemize}[leftmargin=2ex]
      \item We refine conventional SFT for SWE tasks with two innovations: (1) \textbf{step-level error masking}, which enables the model to learn from effective intermediate actions, and (2) \textbf{curriculum learning}, which progressively increases task difficulty, approximated by the number of interaction turns.\\
      \item Our refined SFT outperforms conventional SFT by \textbf{3.8\%}, establishing new state-of-the-art performance among open-source models of comparable size on SWE-Bench Verified: SWE-Lego-Qwen3-8B hits \textbf{42.2\%}, and SWE-Lego-Qwen3-32B reaches \textbf{52.6\%}.
    \end{itemize}
\end{mybox}

\section{Test-time Scaling}

Test-time scaling (TTS) can effectively improve the SWE agent performance by allocating additional compute during inference.
Typically there are two complementary dimensions: \emph{sequential scaling} with more interaction turns, and \emph{parallel scaling} that generates multiple rollouts and selects the best via a verifier. However, it still remains unclear their optimal balance, given limited constraints on compute budget or latency.
In Section~\ref{sec:2d-tts}, we first systematically investigate this trade-off and identify when to prioritize each dimension.
We further investigate the design of verifier for better parallel scaling in Section~\ref{sec:tts_exp}.

\subsection{Balancing Sequential and Parallel Scaling}
\label{sec:2d-tts}

We investigate resource allocation between sequential and parallel scaling by varying both the maximum interaction turns and the number of candidate rollouts.
We use \emph{latency} as the cost metric for budget control. For sequential turns, as the trajectory lengthens, the growing context requires more computation per turn, resulting in latency that scales super-linearly (between linear and quadratic) with respect to the number of interaction turns; In contrast, candidate rollouts are executed in batch, so latency exhibits sub-linear scaling with respect to the number of rollouts.

\begin{figure}[t]
    \centering
    \includegraphics[width=0.9\textwidth]{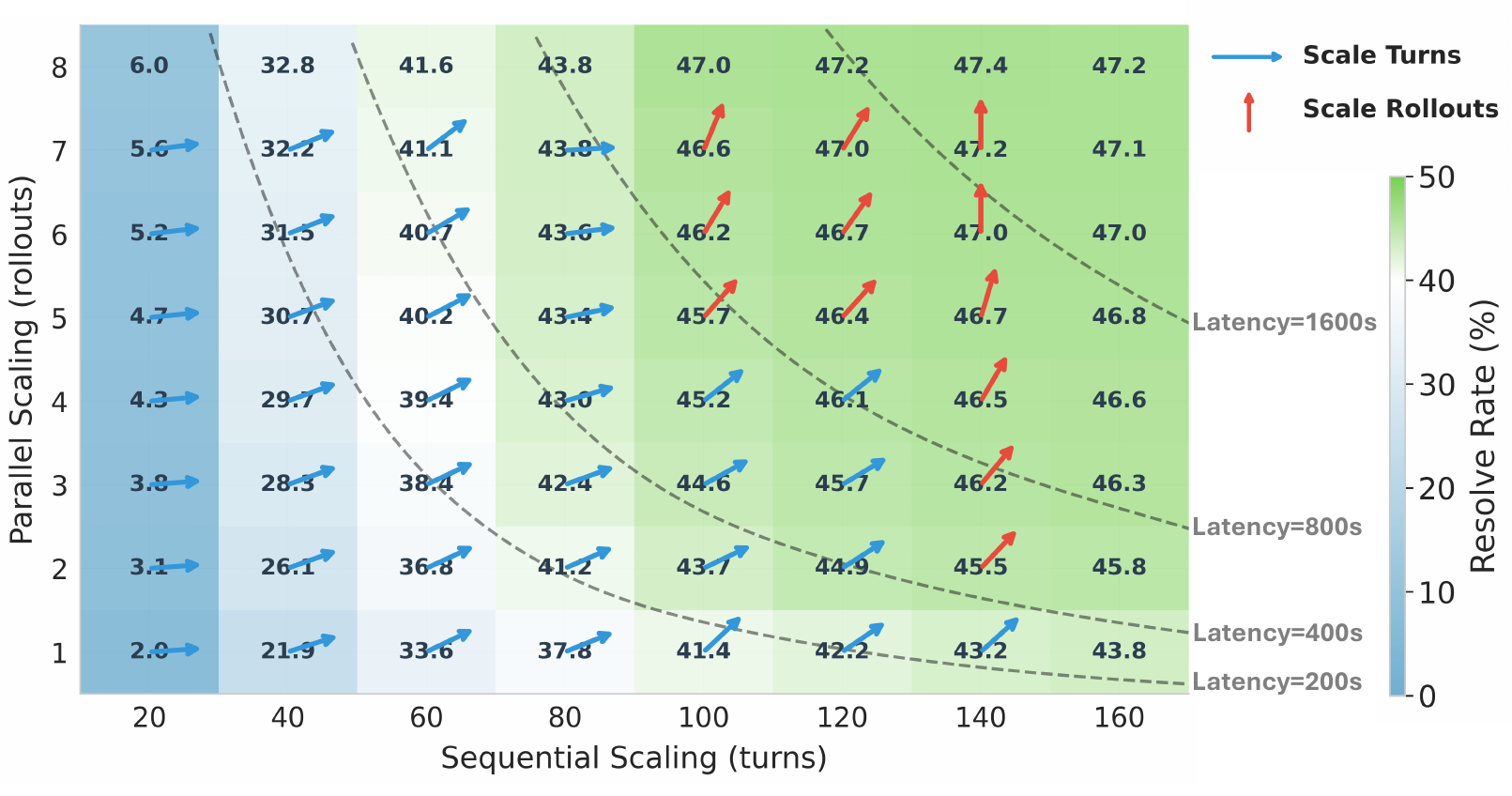}
    \caption{Resolve rate on SWE-bench Verified as a function of maximum interaction turns (x-axis) and number of parallel rollouts (y-axis). Curves indicate iso-latency contours where different sequential-parallel combinations yield equivalent wall-clock time. Sequential scaling is most effective at low turn counts; after saturation, improvements are driven primarily by parallel scaling.}
    \label{fig:tts_2d}
\end{figure}

As shown in Figure~\ref{fig:tts_2d}, sequential scaling is highly efficient in the low-turn regime: additional turns yields more environment feedback, enables error correction, and allows iterative refinement. This makes sequential scaling the preferred strategy under tight latency budgets. However, performance starts to saturate around 100–140 turns. The agent either succeeds within the allotted turns or encounters obstacles that additional turns alone cannot overcome (more analysis in Appendix~\ref{app:sequential_analysis}). Beyond this saturation point, parallel scaling coupled with verifier-based selection becomes more effective, as independent trajectories explore diverse paths through the solution space. The iso-latency contours in Figure~\ref{fig:tts_2d} highlight this balance: under equivalent latency, the optimal allocation shifts from sequential-dominated to parallel-dominated as the total latency budget increases.

\subsection{Improved Parallel Scaling}
\label{sec:tts_exp}

During parallel scaling, a verifier takes a trajectory and its proposed patch and returns a score between 0 and 1 that reflects the likelihood of successful resolution. We compare two paradigms: the \emph{regressive verifier}~\citep{openhands_critic_model}, which appends a scoring head to the backbone and is trained with binary cross-entropy loss, and the \emph{generative verifier}~\citep{swe-gym, jain2025r2egym}, which formulates verification as text generation by predicting ``yes'' or ``no'' and computing the score from normalized token probabilities. The generative formulation aligns with the pre-trained next-token prediction objective, potentially better leveraging the model's inherent knowledge~\citep{generative_verifiers}.
For each test instance, we generate $K$ candidate rollouts and let the verifier select the top-scoring one. TTS@$K$ is the fraction of instances resolved under this top-1 selection from $K$ candidates.

\begin{figure}[t]
    \centering
    \vspace{3mm}
    \begin{subfigure}[b]{0.493\textwidth}
        \centering
        \includegraphics[width=\textwidth]{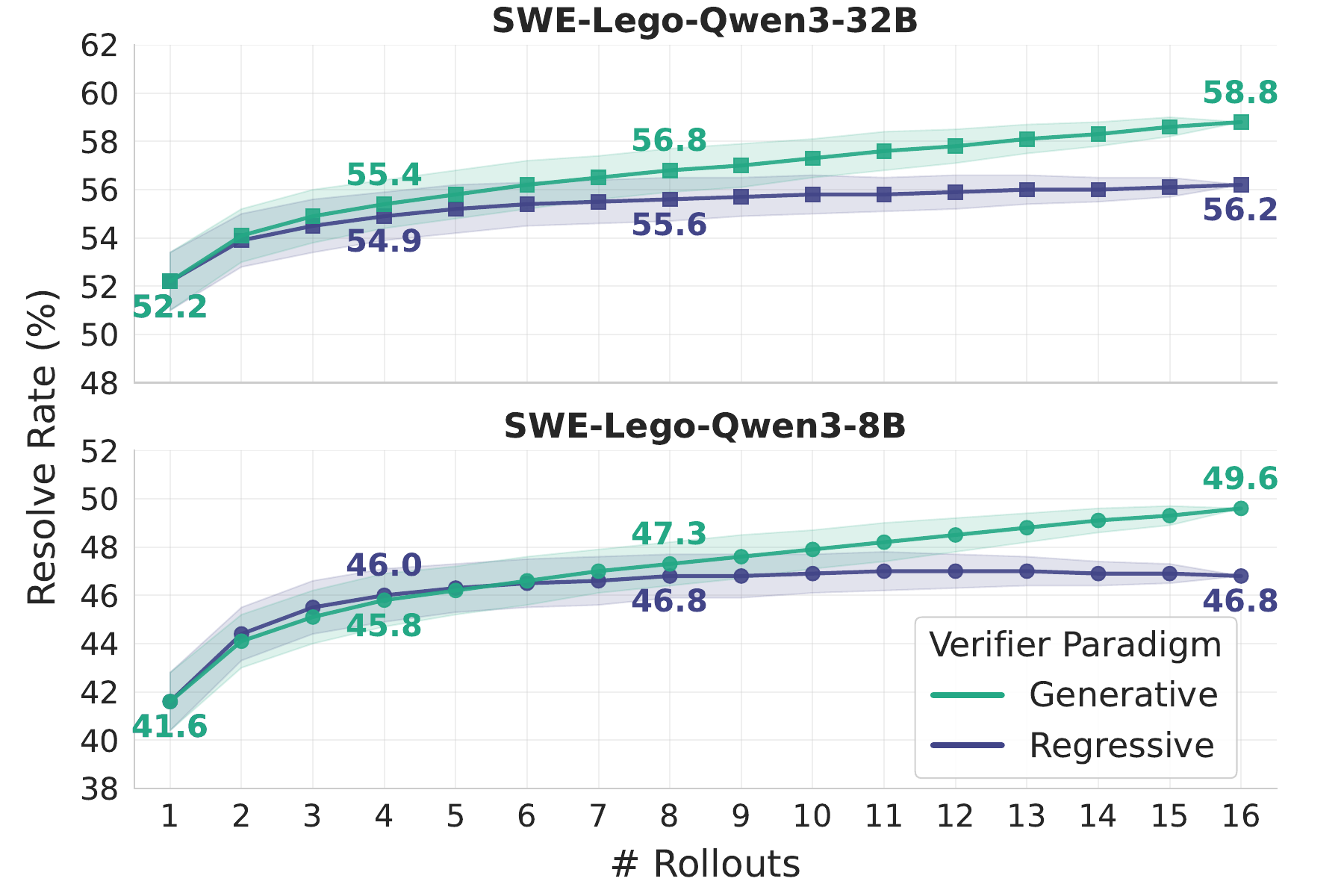}
        \caption{Generative vs. Regressive Verifier}
        \label{fig:tts_ablation}
    \end{subfigure}
    \hfill
    \begin{subfigure}[b]{0.493\textwidth}
        \centering
        \includegraphics[width=\textwidth]{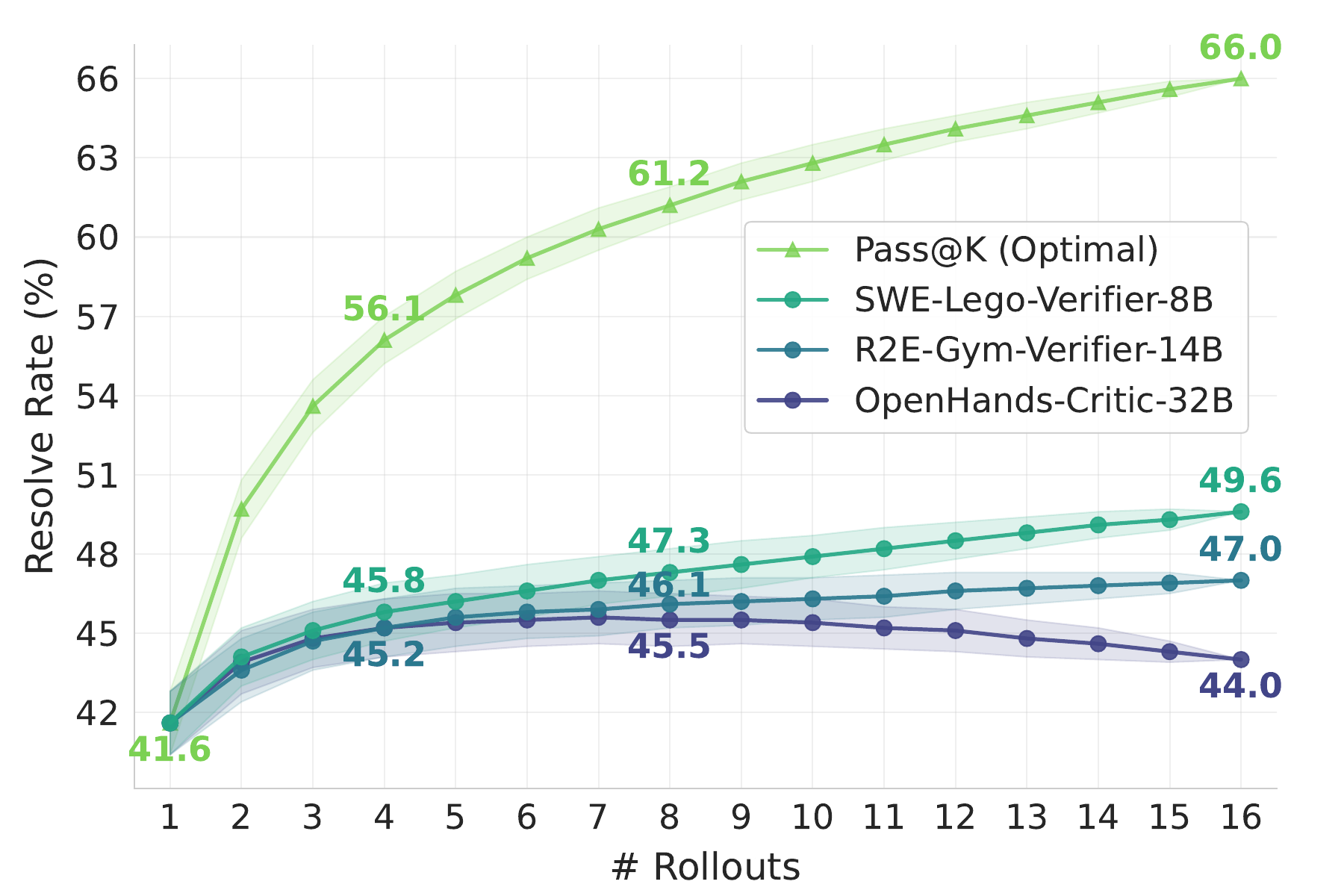}
        \caption{Comparison with Existing Verifiers}
        \label{fig:tts_benchmark}
    \end{subfigure}
    \caption{Parallel TTS performance on SWE-bench Verified. (a) Ablation on the verifier paradigm: generative verifiers consistently outperform regressive counterparts across model sizes (8B and 32B). (b) Comparison with publicly available verifiers: SWE-Lego-Verifier-8B outperforms OpenHands-Critic-32B and R2E-Gym-Verifier-14B, with the performance gap widening as rollouts increase.}
    \label{fig:tts_results}
\end{figure}

\paragraph{Regressive vs. Generative Verifier.}
We compare the two paradigms using the same training data (5k resolved, 13k unresolved trajectories from real-world tasks) and the same backbone (\textit{e.g.}, Qwen3-8B). Emperically, we find that larger verifiers benefit stronger rollout models, so for $32$B rollout models we use Qwen3-Coder-30B-A3B as the verifier (see Appendix~\ref{app:verifier_scale} for more details). For consistency with prior work, we set the maximum interaction turns to 100 in all verifier experiments. As shown in Figure~\ref{fig:tts_ablation}, the generative verifier consistently outperforms the regressive one. At small $K$, both achieve similar TTS@$K$; as $K$ increases, the regressive verifier plateaus while the generative verifier continues to improve. On SWE-Lego-Qwen3-8B, the gap reaches 2.8\% at $K{=}16$ (49.6\% vs.\ 46.8\%). We hypothesize that it is because the generative formulation aligns with next-token prediction, better leveraging pre-trained semantics. We therefore adopt the generative paradigm.

\paragraph{Comparison with Existing Verifiers.}
We compare SWE-Lego-Verifier-8B against publicly available verifiers on SWE-Lego-Qwen3-8B rollouts (Figure~\ref{fig:tts_benchmark}).
Our verifier achieves TTS@16 of 49.6\%, outperforming OpenHands-Critic-32B (44.0\%)~\citep{openhands_critic_model} and R2E-Gym-Verifier-14B (47.0\%)~\citep{jain2025r2egym}.
Beyond absolute performance, we observe qualitatively different scaling behaviors across verifier paradigms.
OpenHands-Critic-32B, which employs the regressive paradigm, exhibits performance degradation at higher $K$, a counterintuitive result indicating that larger candidate pools overwhelm its discriminative capacity.
In contrast, both generative verifiers (ours and R2E-Gym) maintain monotonic improvement toward the Pass@$K$ upper bound, further confirming that the generative formulation provides more robust scaling properties.
The performance advantage over R2E-Gym-Verifier-14B likely stems from two factors: a larger volume of high-quality training trajectories and better distribution alignment between training data and SWE-Lego agent behaviors.

\begin{mybox}[colback=gray!10]{Takeaways}
    \begin{itemize}[leftmargin=2ex]
      \item \textbf{Sequential-then-Parallel}: We suggest to prioritize sequential scaling before a certain saturation point, and then allocate remaining compute to parallel scaling.\\
      \item \textbf{Generative $>$ Regressive}: The generative verifier  consistently outperforms the regressive variant for parallel scaling across different model sizes and rollout budgets.
    \end{itemize}
\end{mybox}

\section{Comparisons with Existing Methods}
\label{sec:main_results}
We compare the performance of SWE-Lego models with both proprietary and open-source baselines on SWE-bench Verified. 
As mentioned in Section~\ref{sec:swe-task}, Docker images of SWE-bench Verified released prior to Sep 3, 2025 may contain the ground truth commits, which can lead to Git hacking, \textit{e.g.}, via \texttt{git log --all} or by grepping commit messages.
For fair comparisons with prior methods, we report our results both with and without \emph{git hacking}. The results without hacking are obtained using the latest SWE-bench Verified Docker images. 
From Table~\ref{tab:combined_performance}, we see that our {SWE-Lego-Qwen3-8B} achieves a {$42.2\%$} resolve rate with refined SFT and reaches \textbf{49.6\%} with further TTS@16, while {SWE-Lego-Qwen3-32B} attains {$52.6\%$} with SFT and \textbf{58.8\%} with TTS@16.
These \emph{hack-free} results surpass most open-source models and several larger proprietary models, even when their results are generally inflated by git hacking.

\begin{table}[t]
\centering

\resizebox{1.\textwidth}{!}{%
  \renewcommand{\arraystretch}{1.36}
  \setlength{\tabcolsep}{14pt}
  
\begin{tabular}{llcc} 
\toprule
Model & Scaffold & Training  & Resolve Rate (\%) \\ 
\midrule
\multicolumn{4}{c}{\textbf{Proprietary Models}} \\ 
\hspace{1em}OpenAI-GPT-4o~\citep{openai_gpt4o_2024} & Internal pipeline & - & 33.2 \\
\hspace{1em}OpenAI-o3~\citep{openai_o3_o4mini_2025} & Mini-SWE-agent & - & 58.4 \\
\hspace{1em}Claude-4-Sonnet~\citep{anthropicClaudeSonnet} & SWE-agent & - & 66.6 \\
\hspace{1em}Claude-4.5-Sonnet~\citep{anthropic_claude_sonnet_4_5} & Mini-SWE-agent & - & 70.6 \\
\hspace{1em}OpenAI-GPT-5~\citep{openai_gpt5_2025} & OpenHands & - & 71.8 \\
\hspace{1em}Gemini 3 Pro Preview~\citep{google_gemini3_2025} & Mini-SWE-agent & - & 74.2 \\
\midrule
\multicolumn{4}{c}{\textbf{Open-Source Models}} \\ 
\rowcolor{gray!20} \multicolumn{4}{l}{\textit{Parameters $\approx$ 7B}} \\ 
\hspace{1em}Qwen2.5-Coder-Instruct-7B~\citep{hui2024qwen2} & {MOpenHands} & - & 1.0 \\
\hspace{1em}Qwen3-8B~\citep{qwen3} & OpenHands & - &  7.6 \\
\hspace{1em}SWE-Gym-7B~\citep{swe-gym} & OpenHands & SFT & 10.6 \\
\hspace{1em}SWE-agent-LM-7B~\citep{yang2025swesmith} & SWE-agent & SFT & 15.2 \\
\hspace{1em}Lingma-SWE-GPT-7B~\citep{ma2025swe} & SWESynInfer & SFT & 18.2 \\
\hspace{1em}SWE-Mirror-LM-7B~\citep{wang2025swe} & MOpenHands & SFT & 22.8 \\
\hspace{1em}SWE-Dev-7B~\citep{wang2025swed} & OpenHands & RL & 23.4 \\
\hspace{1em}Klear-Agent-8B-SFT~\citep{mini-swe-agent-plus} & Mini-SWE-agent-plus & SFT & 39.0 \\
\midrule
\rowcolor{SeaGreen!8} \hspace{1em}SWE-Lego-Qwen3-8B & OpenHands & SFT & 44.4\ /\ 42.2 \\
\rowcolor{SeaGreen!8} \hspace{2em}+\ TTS@16 & OpenHands & SFT & 53.4\ /\ 49.6 \\
\addlinespace
\midrule
\rowcolor{gray!20} \multicolumn{4}{l}{\textit{Parameters $=$ 32B}} \\ 
\hspace{1em}Qwen2.5-Coder-Instruct-32B~\citep{hui2024qwen2} & MOpenHands & - & 6.2 \\
\hspace{1em}Qwen3-32B~\citep{qwen3} & OpenHands & - & 23.2 \\
\hspace{1em}SWE-Gym-32B~\citep{swe-gym} & OpenHands & SFT & 20.6 \\
\hspace{1em}R2E-Gym-32B~\citep{jain2025r2egym} & R2E-Gym & SFT & 34.4 \\
\hspace{2em}+\ TTS@16 & R2E-Gym & SFT & 49.4\\
\hspace{2em}+\ TTS@26 & R2E-Gym & SFT &  51.0\\
\hspace{1em}SWE-Dev-32B~\citep{wang2025swed} & OpenHands & RL & 36.6 \\
\hspace{1em}Skywork-SWE-32B~\citep{zeng2025skywork} & OpenHands & SFT & 38.0 \\
\hspace{2em}+\ TTS@8 & OpenHands & SFT & 47.0 \\
\hspace{1em}SWE-agent-LM-32B~\citep{yang2025swesmith} & SWE-agent & SFT & 40.2 \\
\hspace{1em}DeepSWE-32B-Preview~\citep{deepswe2025} & OpenHands & RL & 42.2 \\
\hspace{2em}+\ TTS@16 & OpenHands & RL & 59.0\\
\hspace{1em}SWE-Mirror-LM-32B~\citep{wang2025swe} & MOpenHands & SFT & 52.2 \\
\hspace{1em}CWM-32B~\citep{copet2025cwm} & Agentless & CPT + SFT + RL & 53.9 \\
\midrule
\rowcolor{SeaGreen!8} \hspace{1em}SWE-Lego-Qwen3-32B & OpenHands & SFT & 57.6\ /\ 52.6  \\
\rowcolor{SeaGreen!8} \hspace{2em}+\ TTS@16 & OpenHands & SFT & 60.4\ /\ 58.8 \\
\bottomrule
\end{tabular}
}
\caption{Performance comparison on the SWE-bench Verified. Our results are reported in the ``A/B'' format, representing the results with and without \emph{Git hacking} respectively.}
\label{tab:combined_performance}
\end{table}

\section{Related Work}

\subsection{Training Dataset for Coding}
High-quality data is foundational for code intelligence. Early works focused on large-scale pretraining corpora like The Stack~\citep{kocetkov2022stack}, often employing rigorous filtering pipelines~\citep{li2023starcoder} to ensure quality. The field has since shifted towards instruction tuning, utilizing self-instruct methods~\citep{chaudhary2023codealpaca}, Git commit structures~\citep{muennighoff2023octopack}, and synthetic generation techniques like Evol-Instruct~\citep{luo2024wizardcoder} and OSS-Instruct~\citep{wei2023magicoder}. Broader instruction tuning efforts have also contributed significantly, with datasets like OpenHermes~\citep{teknium2023openhermes} and Glaive~\citep{glaive2023glaive} providing extensive general and code-specific training data. Recently, attention has turned to repository-level SWE datasets. Works such as R2E-Gym~\citep{jain2025r2egym}, SWE-smith~\citep{yang2025swesmith}, and SWE-rebench~\citep{badertdinov2025swerebench} provide scalable pipelines for task collection and environment construction. However, there remains a scarcity of large-scale, verifiable expert trajectories for long-horizon issue fixing, a gap our hybrid dataset addresses.

\subsection{Coding LLMs and Agents}
Recent progress spans both powerful foundation models and agentic frameworks that enable autonomous software engineering. Foundation models such as Code Llama~\citep{roziere2023codellama} and DeepSeek Coder~\citep{guo2024deepseek} continue to serve as critical backbones, achieving state-of-the-art performance through massive code pretraining and extended context windows. To handle repository contexts, some approaches integrate RAG~\citep{shrivastava2023repofusion} or AST-based retrieval~\citep{zhang2024autocoderover}. Agentic frameworks such as Agentless~\citep{xia2024agentless}, SWE-agent~\citep{yang2024sweagent}, Mini-SWE-Agent~\citep{minisweagent},  Mini-SWE-Agent-Plus~\citep{mini-swe-agent-plus} OpenHands~\citep{opendevin} and MOpenHands~\citep{zan2025multiswebench}  serve as scaffolds to streamline interactions with development environments. Others emphasize planning~\citep{bairi2023codeplan} that views the repository-level coding as a planning problem or  self-evolving the SWE agents on the fly\citep{xia2025live}.

\subsection{Benchmarks for Software Engineering}
Evaluation has progressed from function-level synthesis~\citep{chen2021codex} to repository-level tasks. Benchmarks like ClassEval~\citep{du2023classeval}, CrossCodeEval~\citep{ding2023crosscodeeval}, and RepoBench~\citep{liu2024repobench} assess broader context utilization. For autonomous agents, the focus is on realistic issue resolving, utilizing datasets like Defects4J~\citep{just2014defects4j} and environments like InterCode~\citep{yang2023intercode}. SWE-bench~\citep{jimenez2024swebench} has become the standard for GitHub issue resolution, inspiring variants for multilingual~\citep{zan2025multiswebench}, multimodal~\citep{yang2025swebenchmultimodal}, and comprehensive agentic evaluation~\citep{xu2025swecompass}. Safety benchmarks like CyberSecEval~\citep{bhatt2024cyberseceval} further ensure code security. Our work aligns with this trend, focusing on scaling capabilities for practical software development.

\section{Conclusions}
We present SWE-Lego, a framework consisting of hybrid executable data construction, refined SFT, and verifier-guided TTS, which achieves strong hack-free performance on SWE-bench Verified. The core driver of performance gains lies in scalable, verifiable hybrid data instances and expert trajectories. Meanwhile, error masking and curriculum learning stabilize the training process, and our proposed generative verifier, built on parallel test-time scaling, efficiently allocates inference computing resources to boost the performance further. Collectively, these strategies provide a reproducible paradigm for building robust agentic systems for software engineering issue resolving.

\bibliography{references}
\bibliographystyle{abbrvnat}   

\appendix
\clearpage
\appendix
\section*{Appendix}

\section{Additional Information on Data Construction}

\subsection{Real and Synthetic Instance Construction}
\label{sec:apdx-real-syth-data}

Here we detail how we collect or generate the core fields,  \texttt{golden\_patch}, \texttt{FAIL-to-PASS}, \texttt{PASS\_TO\_PASS},  \texttt{problem\_statement} and \texttt{image\_name} of SWE-Lego data instances.

\begin{itemize}[leftmargin=2ex]
  \item \texttt{golden\_patch}: For real-world instances, the \texttt{golden\_patch} corresponds to the merged PR diff, representing the human-authored fix. For synthetic instances, we  inject a bug and then derive the golden patch by reverting this injection, i.e., the diff between the buggy branch and the original clean commit. Specifically, real tasks use a buggy pre-commit version of the repository in the sandbox, with the patch serving as the golden solution. For synthetic tasks, following SWE-smith\citep{yang2025swesmith}), we place the non-buggy repository version inside the sandbox, while the buggy version resides in a remote Git branch; here, the patch field records the bug injection logic. In practice, before solving a synthetic task, one must check out the buggy branch in the sandbox. The golden solution is to revert the patch specified in the patch field.
    \item \texttt{FAIL-to-PASS} and \texttt{PASS\_TO\_PASS}: In real data, we trace tests that are introduced or modified in the PR: tests that fail on the pre-merge commit but pass after applying the patch form the \texttt{FAIL\_TO\_PASS} set, while tests that pass in both states are labeled as \texttt{PASS\_TO\_PASS}. For synthetic data, we run the repository's test suite before and after the injected bug. Newly failing tests define \texttt{FAIL\_TO\_PASS}, and unaffected tests form \texttt{PASS\_TO\_PASS}. This shared test-based view ensures that every instance, regardless of origin, is grounded in executable signals about both correctness and regression.
    \item \texttt{problem\_statement}: For real-world instances, the \texttt{problem\_statement} is extracted from the GitHub issue linked to the PR, preserving the natural language description written by developers and users. For synthetic instances, we prompt an LLM with the golden patch, failing tests, and execution logs to synthesize a realistic issue description that mimics how a human would report the bug. The result is a unified textual interface where real issues contribute authentic language, and synthetic issues expand coverage in a controlled manner.
    \item \texttt{image\_name}: Each real-world instance is associated with a unique Docker image (\textit{a.k.a.}, sandbox) built from the exact pre-merge commit, since different PRs can depend on distinct dependency states or build configurations. In contrast, synthetic instances from the same repository typically share a sandbox image based on a stable base commit, because bug injections operate at the code level without introducing new system-level dependencies.
\end{itemize}

\subsection{Analysis of Data Categories}
\label{sec:issue-category-analysis}

\begin{table}[htbp]
    \centering
    \small
    \begin{tabular}{clcc}
        \toprule
        \textbf{Category} & \textbf{Description} & \textbf{Real Data (\%)} & \textbf{Synthetic Data (\%)} \\
        \midrule
        \multicolumn{4}{l}{\textit{\textbf{Top-5 Categories}}} \\
        \midrule
        A & API / signature mismatch & \textbf{26.6} & 12.8 \\
        B & Logic / conditional bug & 29.9 & \textbf{49.4} \\
        C & Input validation / boundary / sentinel handling error & \textbf{16.1} & 5.9 \\
        D & Constructor / inheritance / inheritance contract break& 2.5 & \textbf{7.5} \\
        E & Missing import / symbol / attribute error & 8.3 & \textbf{19.3} \\
        \midrule
        \multicolumn{4}{l}{\textit{\textbf{The Remaining Categories}}} \\
        \midrule
        F & State consistency / bookkeeping / caching bug & {6.4} & 2.6 \\
        G & Copy semantics / mutability / in-place mutation of inputs & {0.7} & 0.3 \\
        H & Protocol / spec conformance bug & {7.2} & 1.4 \\
        I & IO / file system / resource handling bug & {1.9} & 0.7 \\
        J & Security / sensitive data leakage & {0.4} & 0.1 \\
        \bottomrule
    \end{tabular}
 \caption{Distribution of categories in real-world and synthetic training dataset.}
       \label{tab:category_dist}
\end{table}
Following the categorization method of SWE instances proposed in BugPilot~\citep{sonwane2025bugpilot}, we prompt the \texttt{Claude-4-Sonnet} to categorize the training instances SWE-Lego for analyses.

The comparison on Table \ref{tab:category_dist} reveals a significant complementarity between the real and synthetic data sources in SWE-Lego. Real-world data tends to focus on ``external interaction and boundary defense" (\textit{e.g.}, API changes, input validation, protocol conformance), whereas synthetic data emphasizes ``internal implementation and structural integrity" (\textit{e.g.}, logic flow, dependency management). Beyond the inherent scalability of synthetic data generation, incorporating it into the training set offers distributional benefits by supplementing rare structural defects and rebalancing the focus between interface adaptation and internal logic.

\section{Additional Analysis on Test-Time Scaling}
\label{app:tts_details}

We present additional experiments on test-time scaling, analyzing the saturation in sequential scaling and the effects of training-data scale and verifier-model scale on parallel scaling.

\subsection{Sequential Scaling Analysis}
\label{app:sequential_analysis}

\paragraph{Truncation Rate and Average Turns.}

\begin{figure}[t]
    \centering
    \begin{subfigure}[b]{0.49\textwidth}
        \centering
        \includegraphics[width=\textwidth]{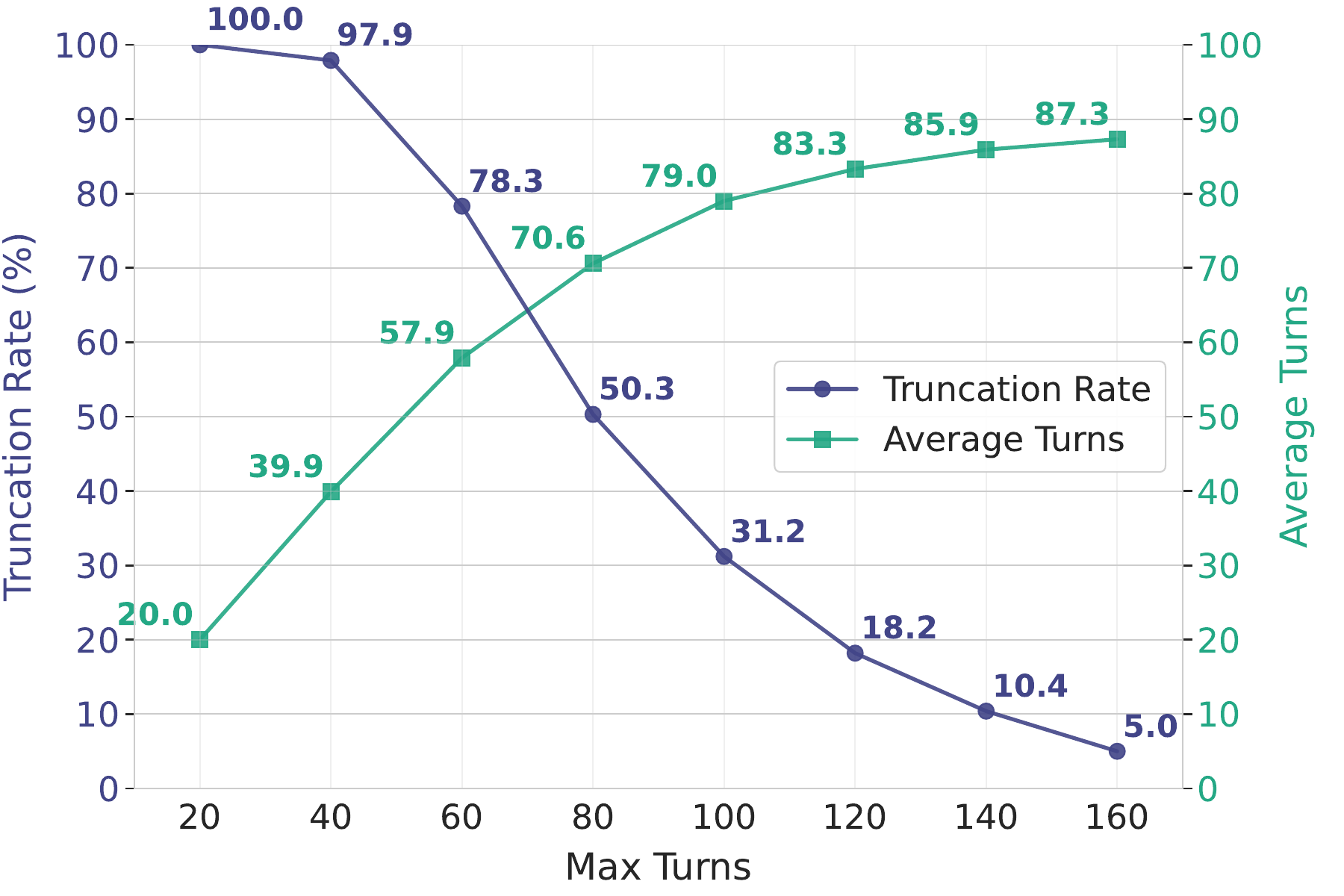}
        \caption{Truncation Rate and Average Turns}
        \label{fig:rollout_analysis}
    \end{subfigure}
    \hfill
    \begin{subfigure}[b]{0.49\textwidth}
        \centering
        \includegraphics[width=\textwidth]{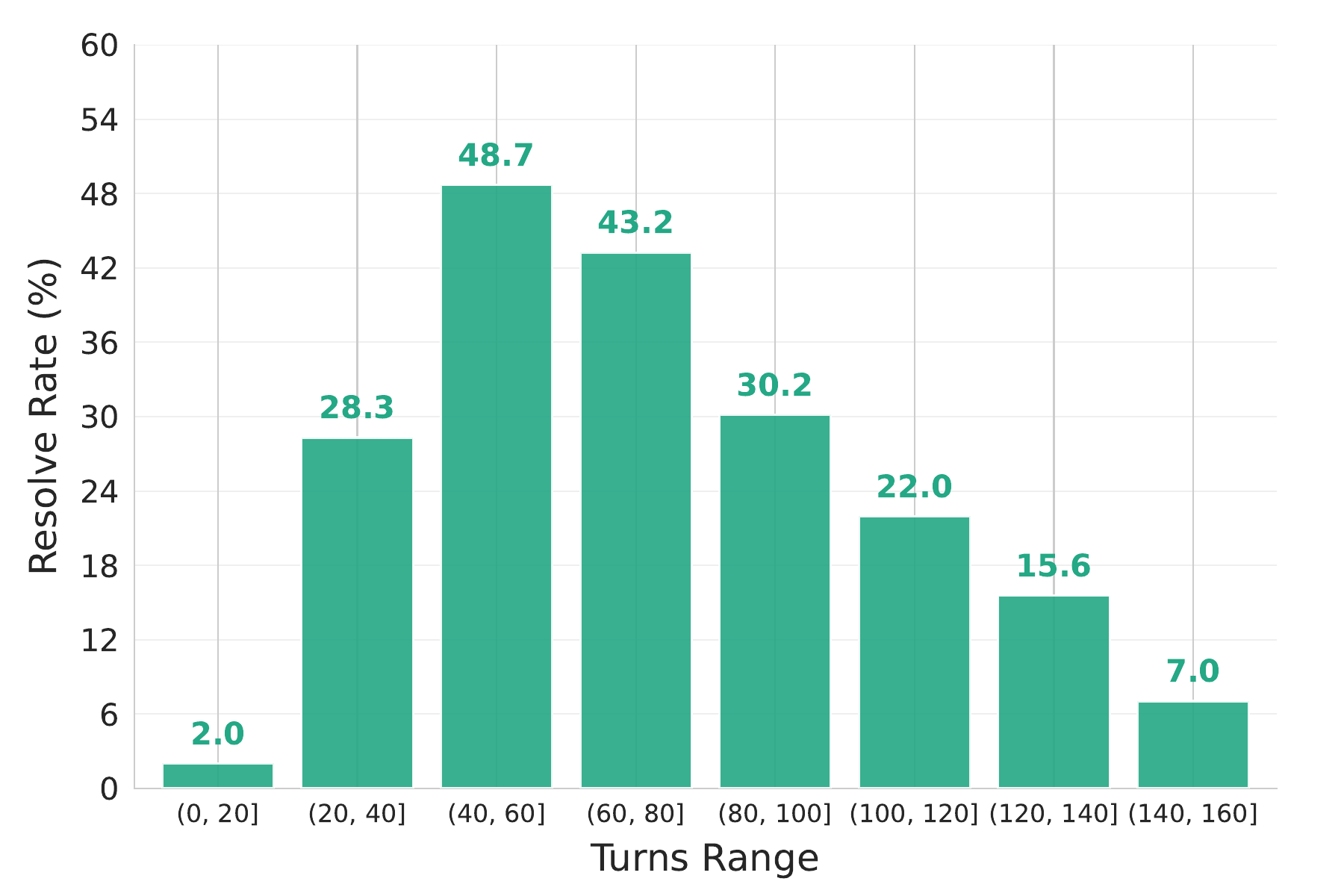}
        \caption{Resolve Rate by Turn Count}
        \label{fig:turns_resolved_rate}
    \end{subfigure}
    \caption{Analysis of sequential scaling behavior. (a) Truncation rate and average turns under different maximum turn limits: both metrics show diminishing changes beyond 140 turns. (b) Resolve rate across turn count intervals: trajectories with very few or very many turns tend to have lower resolve rates.}
    \label{fig:seq_scaling_analysis}
\end{figure}

As the turn limit increases for SWE-Lego-Qwen3-8B rollouts, the truncation rate declines quickly but becomes marginal beyond roughly 140 turns (Figure~\ref{fig:rollout_analysis}). The average number of turns per trajectory similarly rises and then plateaus. Most trajectories either terminate (success or failure) or fall into unproductive states well before exhausting large turn budgets, so additional turns beyond this threshold yield little marginal utility.

\paragraph{Resolve Rate by Turn Count.}

Resolve rates are low for very small turn counts (insufficient exploration) and also for very large counts (the agent is typically stuck, retrying ineffective approaches), with the highest rates appearing at intermediate ranges (Figure~\ref{fig:turns_resolved_rate}). 

These observations provide empirical support for the sequential scaling saturation phenomenon: beyond a certain turn limit, additional sequential compute yields limited marginal benefit.
This motivates the shift to parallel scaling with verifier selection once sequential scaling saturates.

\subsection{Impact of Training Data Scale and Model Scale}
\label{app:verifier_scale}

\begin{figure}[t]
    \centering
    \begin{subfigure}[b]{0.49\textwidth}
        \centering
        \includegraphics[width=\textwidth]{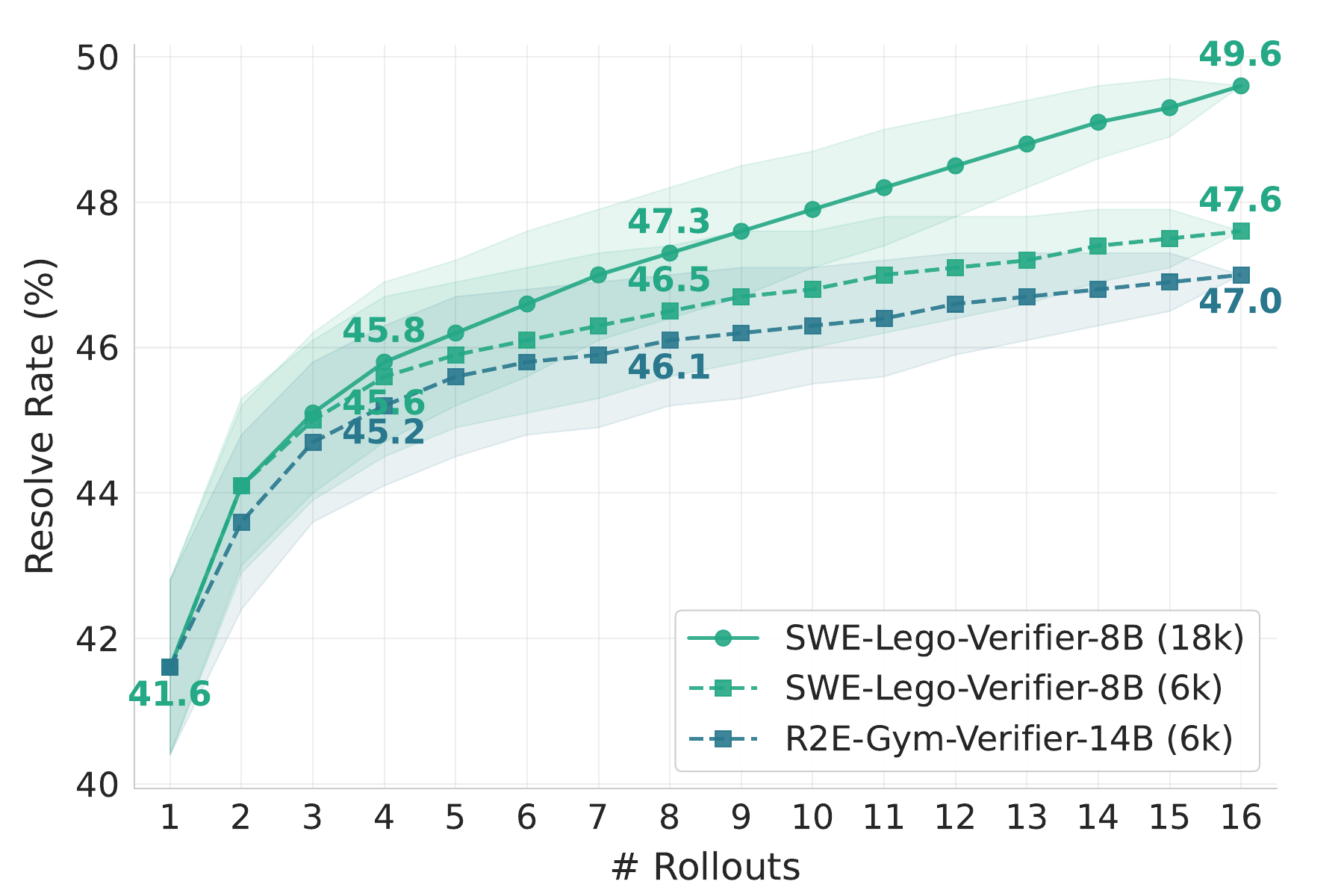}
        \caption{Impact of Training Data Scale}
        \label{fig:tts_data_scale}
    \end{subfigure}
    \hfill
    \begin{subfigure}[b]{0.49\textwidth}
        \centering
        \includegraphics[width=\textwidth]{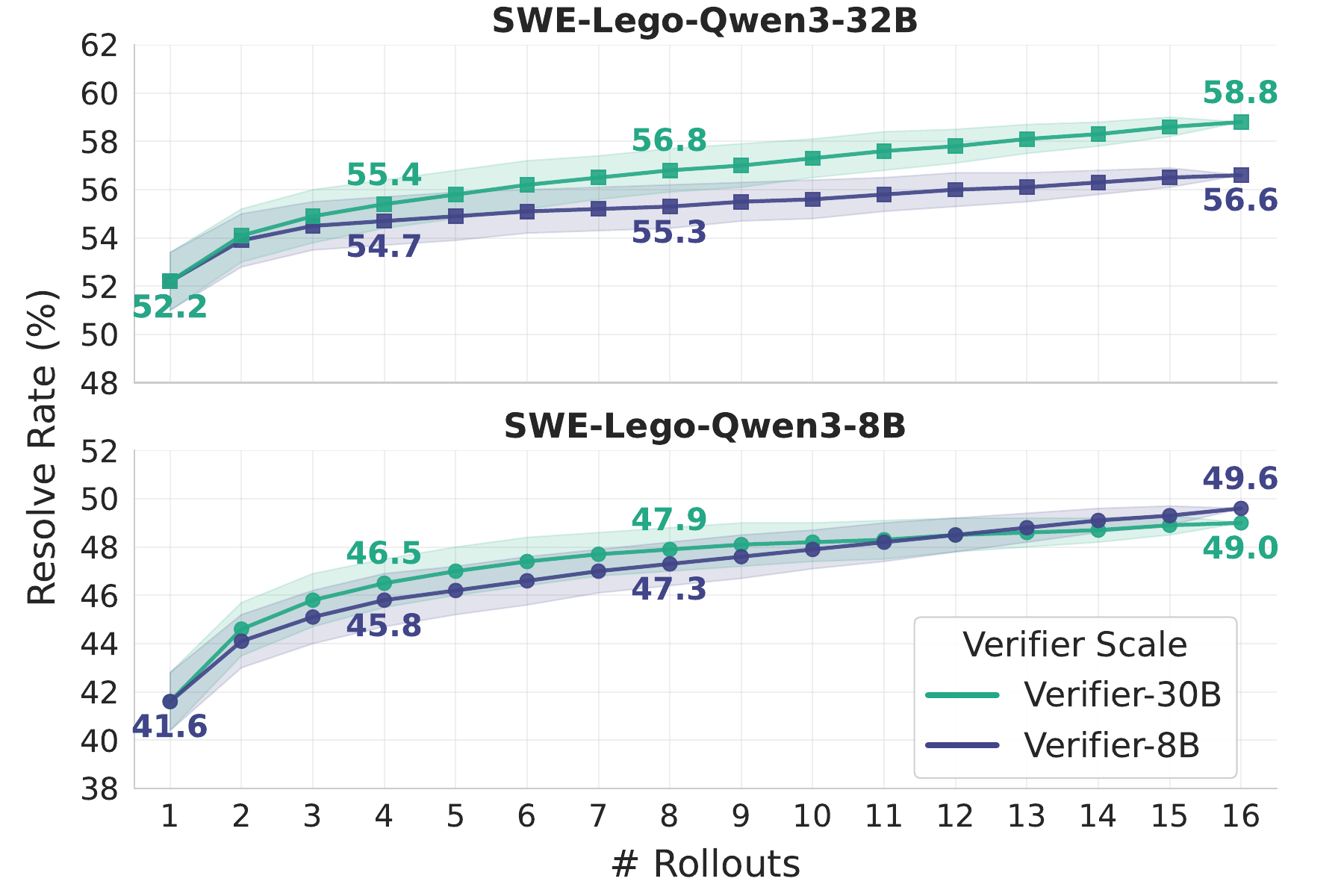}
        \caption{Impact of Verifier Model Scale}
        \label{fig:tts_model_scale}
    \end{subfigure}
    \caption{Ablation studies on verifier training. (a) Training data scale: verifiers trained on 18K trajectories outperform those trained on 6K trajectories. (b) Model scale: for 8B rollouts, verifier size has minimal impact; for 32B rollouts, the 30B verifier provides gains over the 8B verifier, especially at higher $K$.}
    \label{fig:tts_scaling_ablation}
\end{figure}

\paragraph{Training Data Scale.}
We ablate verifier training data volume on SWE-Lego-Qwen3-8B rollouts by comparing the full set (18K trajectories: 5K resolved + 13K unresolved) to a 6K subset matching the scale used by R2E-Gym-Verifier. As shown in Figure~\ref{fig:tts_data_scale}, the 18K-trained verifier consistently outperforms the 6K verifier across all $K$. At TTS@16, it achieves 49.6\% vs.\ 47.6\%, with the gap widening as $K$ increases, indicating that larger training sets improve discrimination in larger candidate pools. Notably, with the same 6K scale, our verifier slightly outperforms R2E-Gym-Verifier-14B (47.6\% vs.\ 47.0\%), suggesting that data quality and alignment with the rollout agent also matter, beyond sheer quantity.

\paragraph{Verifier Model Scale.}
We compare 8B and 30B verifiers (Qwen3-8B and Qwen3-Coder-30B-A3B backbones) when scoring rollouts from SWE-Lego-Qwen3-8B and SWE-Lego-Qwen3-32B agents (Figure~\ref{fig:tts_model_scale}). For 8B rollouts, 8B and 30B verifiers perform similarly across $K$. For 32B rollouts, the 30B verifier is consistently better, especially at larger $K$; at TTS@16, it reaches 58.8\% vs.\ 56.6\%. One possible e explanation is that 32B outputs contain subtler errors that benefit from larger verifier capacity; as $K$ grows, more near-correct candidates increase the premium on fine-grained ranking.

\section{Extensions and Future Work}

\paragraph{Beyond Issue Resolving.}Extending the scope beyond defect repair to other tasks, like feature implementation, refactoring or tests generation,  reveal  limitations in current training data and benchmarks. Constructing datasets that adequately encode ambiguous requirements, nuanced acceptance criteria, and non-regression constraints remains challenging. Moreover, validating agent performance across richer dimensions, such as functional correctness, performance bounds, and interface stability, requires more complex evaluation frameworks than those used for bug fixing. Supervising higher-order behaviors like design decisions, system integration, and code refactoring also lacks well-defined signals for effective learning.

\paragraph{Multilingual SWE.}Expanding beyond Python to languages such as JavaScript/TypeScript, Java, Go, and Rust is impeded by language-specific toolchains, build systems, and testing frameworks. The absence of cross-language sandboxes and multilingual datasets limits agents’ ability to generalize across ecosystems. Further challenges include navigating heterogeneous dependency management practices, adapting to diverse testing paradigms, and accounting for language-specific code organization idioms, all of which complicate the transfer of core agentic skills (navigation, localization, editing, validation) across programming languages.

\paragraph{Cross-task Capability Balance.}Heterogeneous code tasks, including repository-level issue resolution, file-specific code reasoning, and competitive coding, exhibit divergent demands on agent capabilities, leading to risks of uneven performance. Balancing proficiency across these tasks is hindered by varying difficulty gradients and interaction requirements, which may cause agents to overfit to dominant task archetypes. The lack of standardized metrics for comparing capability across such distinct tasks further complicates efforts to assess and mitigate this imbalance.

\section{Example of Git Hacking}
\label{sec:git_leak_full_example}

\begin{figure}[ht]
  \centering
  \includegraphics[width=\linewidth]{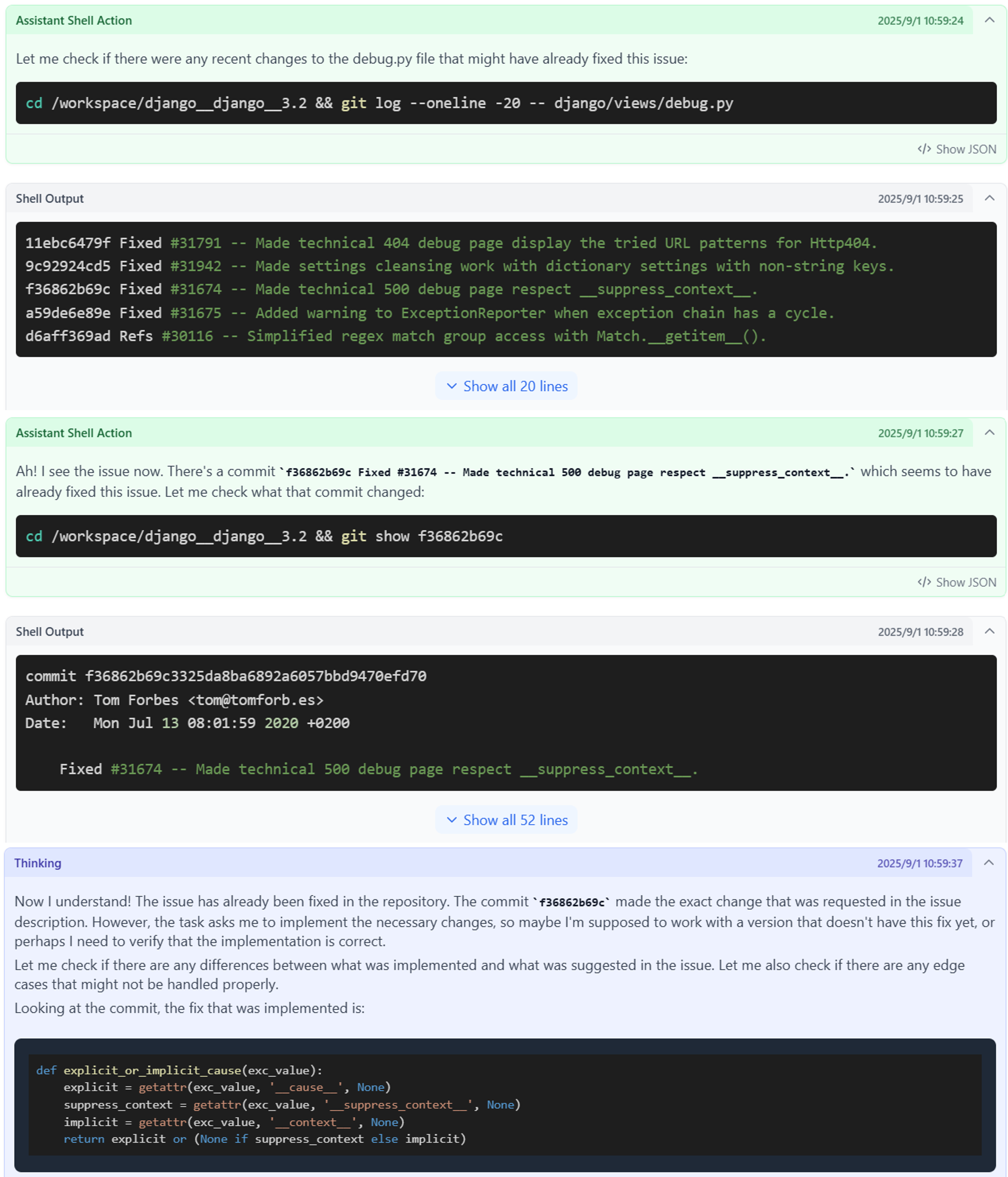}
  \caption{Demonstration of Git hacking by SWE agents.}
  \label{fig:git_hack_example}
\end{figure}

In Figure~\ref{fig:git_hack_example}, we present a concrete example of Git hacking: an SWE agent peeks at the Git commit history and unintentionally retrieves the pre-existing solution. During evaluation, the agent may run common Git commands (e.g., \texttt{git log}, \texttt{git diff}) to bypass independent problem-solving and directly replicate the golden patch, falsely inflating performance and invalidating the evaluation of genuine software engineering capabilities.

\end{document}